\documentclass[fleqn,usenatbib]{mnras}

\usepackage{hyperref}
\usepackage{siunitx}
\usepackage{amsmath}
\usepackage{booktabs}
\let\oldAA\AA
\renewcommand{\AA}{\text{\normalfont\oldAA}}
\usepackage[T1]{fontenc}
\usepackage{xcolor}
\DeclareRobustCommand{\VAN}[3]{#2}
\let\VANthebibliography\thebibliography
\def\thebibliography{\DeclareRobustCommand{\VAN}[3]{##3}\VANthebibliography}

\usepackage{graphicx}
\usepackage{grffile}% Including figure files
\usepackage{newtxtext,newtxmath}

\title[The cosmic merger rate density of compact objects]%{The cosmic merger rate density of compact objects {\filippo{objects or binaries???}}: impact of star formation, metallicity, initial mass function and binary evolution {\filippo{... and the imf?}} {\filippo{title change: what about "A thorough exploration of the parameter space that affects the cosmic merger rate density of compact binaries" ? }}}
{The cosmic merger rate density of compact objects: impact of star formation, metallicity, initial mass function and binary evolution}

\author[Santoliquido et al.]{Filippo Santoliquido,$^{1,2}$\thanks{E-mail: \href{filippo.santoliquido@phd.unipd.it}{filippo.santoliquido@phd.unipd.it}}
Michela Mapelli,$^{1,2,3}$
Nicola Giacobbo,$^{1,2,3}$
\newauthor
Yann Bouffanais,$^{1,2}$
M. Celeste Artale$^{4}$
\\
$^{1}$Dipartimento di Fisica e Astronomia 'G. Galilei', Università degli studi di Padova, vicolo dell'Osservatorio 3, I-35122, Padova, Italia\\
$^{2}$INFN, Sezione di Padova, via Marzolo 8, I-35131, Padova, Italia\\
$^{3}$INAF, Osservatorio Astronomico di Padova, vicolo dell'Osservatorio 5, I-35122, Padova, Italia \\
$^{4}$Institut f{\"u}r  Astro- und Teilchenphysik, Universit{\"a}t Innsbruck, Technikerstrasse 25/8, A-6020, Innsbruck, {\"O}sterreich
}
\date{Accepted XXX. Received YYY; in original form ZZZ}

\pubyear{2020}

\begin{document}
\label{firstpage}
\pagerange{\pageref{firstpage}--\pageref{lastpage}}
\maketitle

\begin{abstract}
We evaluate the redshift distribution of binary black hole (BBH), black hole -- neutron star binary (BHNS) and binary neutron star (BNS) mergers, exploring the main sources of uncertainty: star formation rate (SFR) density, metallicity evolution, common envelope, mass transfer via Roche lobe overflow, natal kicks, core-collapse supernova model and initial mass function. Among binary evolution processes, uncertainties on common envelope ejection have a major impact: the local merger rate density of BNSs varies from $\sim{}10^3$ to $\sim{}20$ Gpc$^{-3}$ yr$^{-1}$ if we change the common envelope efficiency parameter  from $\alpha_{\rm CE}=7$ to 0.5, while the local merger rates of BBHs and BHNSs vary by a factor of $\sim{}2-3$. The BBH merger rate changes by one order of magnitude, when $1 \sigma$ uncertainties on metallicity evolution are taken into account. In contrast, the BNS merger rate is almost insensitive to metallicity. Hence, BNSs are the ideal test bed to put constraints on uncertain binary evolution processes, such as common envelope and natal kicks.  Only models assuming values of $\alpha_{\rm CE}\gtrsim{}2$   and moderately low natal kicks (depending on the ejected mass and the SN mechanism), result in a local BNS merger rate density within the 90\% credible interval inferred from the second gravitational-wave transient catalogue.
%a one-dimensional root mean square velocity $=265 $ km s$^{-1}$, result in a local BNS merger rate density below  the 90\% credible interval inferred from O1, O2 and O3a.}
\end{abstract}


%%%%%%%%%% BODY OF PAPER%%%%%
\section{Introduction}
Gravitational-wave (GW) observations give us an insight into the merger rate density of binary compact objects in the local Universe \citep{abadie2010,abbottGW150914,abbottastrophysics,abbottO1,abbottO2,abbottO2popandrate}. Based on the results of the first (O1) and the second observing runs (O2), the LIGO-Virgo collaboration (LVC) has inferred a local merger rate density $\mathcal{R}_{\rm BBH}\sim{}24-140~\text{Gpc}^{-3}~\text{yr}^{-1}$ and $\mathcal{R}_{\rm BHNS}<610~\text{Gpc}^{-3}~\text{yr}^{-1}$ for binary black holes (BBHs) and black hole -- neutron star binaries (BHNSs), respectively \citep{abbottO2,abbottO2popandrate}.

While this paper was in the review stage, the LVC  published 39  events observed during the first half of the third observing run (O3a, \citealt{abbottO3a}). This leads to a sample of 50 binary compact object mergers from O1, O2 and O3a, known as the second GW transient catalogue (GWTC-2). From these new data and assuming the {\sc{power law + peak}} mass distribution model (which is shown to be preferred by the data), the BBH merger rate density inside the 90\% credible interval is estimated to be $\mathcal{R}_{\rm{BBH}} = 23.9^{+14.9}_{-8.6}$ Gpc$^{-3}$ yr$^{-1}$    ($\mathcal{R}_{\rm{BBH}} = 58^{+54}_{-29}$ Gpc$^{-3}$ yr$^{-1}$) if we exclude (include) the event GW190814 \citep{abbottpopO3a}. 
%This merger rate density has been calculated including  only binary systems with $m_1 \geq m_2 \geq 3 M_\odot$ in the analysis; in other words, GW190814-like events \citep{abbottGW190814} are excluded. If systems down to $m_2 \geq 2 M_\odot $ are also taken into account, the LVC infers a merger rate density  $\mathcal{R}_{\rm{BBH}} = 58^{+54}_{-29}$ Gpc$^{-3}$ yr$^{-1}$ within the 90\% credible interval, always assuming the {\sc{Power law + Peak}} mass distribution model.}
 \cite{abbottGW190425} inferred a BNS local merger rate density $\mathcal{R}_{\rm BNS}=250-2810~\text{Gpc}^{-3}~\text{yr}^{-1}$ from the two published binary neutron star (BNS) mergers, GW170817 \citep{abbottGW170817,abbottmultimessenger} and GW190425 \citep{abbottGW190425}. This number has recently been revised to account for the entire O3a data, leading to a new estimate $\mathcal{R}_{\rm{BNS}} = 320^{+490}_{-240}$ Gpc$^{-3}$ yr$^{-1}$ \citep{abbottpopO3a}.

%At the time of writing this paper, the LIGO-Virgo collaboration (LVC) has published 13 merging compact binaries (CBs), 3 from the first observing run (O1, \citealt{abbottGW150914,abbottastrophysics,abbottO1}), 8 from the second run (O2, \citealt{abbottO2,abbottO2popandrate}) and 3 new events from the third run (O3, \citealt{abbottGW190425,abbottGW190412,abbottGW190814}). From the events observed in O1 and O2 plus GW190425, the LVC has inferred a local merger rate density  $\mathcal{R}_{\rm BBH}\sim{}24-140~\text{Gpc}^{-3}~\text{yr}^{-1}$, $\mathcal{R}_{\rm BHNS}<610~\text{Gpc}^{-3}~\text{yr}^{-1}$ and   $\mathcal{R}_{\rm BNS}=250-2810~\text{Gpc}^{-3}~\text{yr}^{-1}$ for binary black holes (BBHs), black holes - neutron stars (BHNSs) and binary neutron stars (BNSs), respectively \citep{abbottO2,abbottO2popandrate,abbottGW190425}.

%%%%%%%%%%%%%%%%%%%%%%%%%%%%%%%%%%%%%%%%%%%%%%%
%Besides GW190425, the LVC has published two additional events from the third observing run (O3): GW190412 \citep{abbottGW190412} and GW190814 \citep{abbottGW190814}, while  more than 50 public event candidates are currently being analysed\footnote{\url{https://gracedb.ligo.org/}}. Hence, we expect that tens of new events will be published as a result of O3,  further constraining the merger rate density as a function of redshift \citep{Fishbach2018,vitale2019,callister2020}. %%The next future will also witness Advanced LIGO, which at design sensitivity will be able to observe two merging black holes (BHs) with masses $(30 + 30)\text{M}_\odot$ out to a cosmological redshift equal to $z \sim{} 1$ \citep{abbottO2}.

Moreover, the target sensitivity of third-generation ground-based GW interferometers, namely the Einstein Telescope in Europe and Cosmic Explorer in the US, will allow us to observe BBH mergers up to $z\gtrsim{}10$ and BNS mergers up to $z\gtrsim{}2$ \citep{punturo2010,reitze2019,kalogera2019,maggiore2020}. This will make possible to fully reconstruct the evolution of the merger rate with redshift, opening new perspectives on the study of binary compact objects. %It is evident that the observatory network will soon search for mergers in the entire Universe and 

From a theoretical perspective, several studies attempt to predict the cosmic merger rate evolution, based on either cosmological simulations \citep{lamberts2016,lamberts2018,oshaughnessy2017,schneider2017,mapelli2017,mapelli2018,mapelli2018b,mapelli2019,toffano2019,artale2019,artale2020,graziani2020} or semi-analytical models \citep{oshaughnessy2010,dominik2013,dominik2015,belczynski2016,eldridge2016,giacobbo2018b,giacobbo2020,boco2019,eldridge2019,baibhav2019,neijssel2019,vitale2019,tang2020}.

Overall, our current understanding of the merger rate evolution is hampered by large uncertainties. On the one hand,  our knowledge of the cosmic star formation rate (SFR)   \citep[e.g.,][]{madaudickinson2014,madau2017}, and the metallicity evolution of stars \citep[e.g.][]{maiolino2008,rafelski2012,madaudickinson2014,maiolinomannucci2018,decia2018,chruslinska2019,chruslinska2020} are affected by a number of observational uncertainties. On the other hand, the very process of binary compact object formation is still matter of debate (see, e.g. \citealt{mandel2018} and \citealt{mapelli2018c} for two recent reviews).

Several formation channels have been proposed for binary compact objects:  binary evolution via common envelope \citep[e.g.,][]{tutukov1973,bethe1998,portegieszwart1998,belczynski2002,belczynski2008,voss2003,podsiadlowski2004,belczynski2016,eldridge2016,stevenson2017,giacobbo2018b,kruckow2018,vignagomez2018,spera2019,tanikawa2020} or via chemical mixing \citep[e.g.,][]{marchant2016,demink2016,mandel2016}, dynamical evolution in triples  \citep[e.g.,][]{antonini2016,antonini2017,arcasedda2018,fragione2019}, in young star clusters \citep[e.g.,][]{banerjee2010,ziosi2014,mapelli2016,banerjee2017,banerjee2020,kumamoto2019,dicarlo2019a,dicarlo2020,rastello2020}, in globular clusters \citep[e.g.,][]{portegieszwart2000,downing2010,rodriguez2015,rodriguez2016,rodriguez2018,samsing2014,askar2017,samsing2018,fragionekocsis2018,zevin2019,fragione2020,antonini2020} and in galactic nuclei \citep[e.g.,][]{oleary2009,miller2009,mckernan2012,mckernan2018,antonini2016,bartos2017,stone2017,rasskazov2019,arcasedda2020,arcasedda2020b,yang2019,tagawa2020}. Each of these formation channels will likely leave an imprint on the evolution of the merger rate density with redshift (e.g. \citealt{dominik2013,dominik2015,mapelli2017,mapelli2018,rodriguezloeb2018,choksi2018,choksi2019,eldridge2019,yang2019,kumamoto2020,dubuisson2020,mapelli2020b,santoliquido2020}). In particular, the merger rate of BBHs can be dramatically affected by dynamics, because of BH masses favouring dynamical exchanges \citep{hills1980}.  %In particular, \cite{santoliquido2020} highlighted the differences between the merger rate evolution of binary compact objects formed in isolation and in young star clusters. 

Even if we restrict our attention to just one possible formation channel, we are faced with major uncertainties. For example, we do not have a satisfactory picture of the process of common envelope. Most population-synthesis models describe it through a free parameter, $\alpha_{\rm CE}$, which was originally meant to indicate the fraction of orbital energy that is transferred to the envelope \citep{webbink1984}. According to its original definition, $\alpha_{\rm CE}$ should assume only values $\le{}1$ and still theoretical models suggest that values of $\alpha_{\rm CE}>1$ better describe the formation of BNSs \citep[e.g.,][]{mapelli2018,fragos2019,giacobbo2020}. %The magnitude of natal kicks is ano (e.g. \citealt{janka2012,bray2016,bray2018,tauris2017,giacobbo2020}), mass transfer efficiency \citep[e.g.,][]{}

Here, we focus on the formation of binary compact objects in isolation, through common envelope evolution, and we investigate all the main sources of uncertainty that affect the merger rate density evolution. In particular, we account for uncertainties on the cosmic SFR, metallicity evolution, common envelope (by varying the $\alpha_{\rm CE}$ parameter over more than one order of magnitude), natal kicks, core-collapse supernova (SN) models, mass transfer efficiency and on the slope of the initial mass function. 
%; and as a result the catalogues of merging compact binaries were obtained with our population-synthesis code {\sc mobse} \cite{giacobbo2018}. 
We use {\sc cosmo$\mathcal{R}$ate} \citep{santoliquido2020}, a semi-analytic code that combines information on cosmic SFR and  metallicity evolution with catalogues of binary compact objects obtained via binary population-synthesis. {\sc cosmo$\mathcal{R}$ate} is computationally optimised to extensively probe the parameter space. 
% and the main uncertainties that affect the cosmic merger rate density. In particular, we assess the impact of various parameters that drive the binary evolution processes, namely the common envelope (CE) ejection efficiency $\alpha_{\rm CE}$, SN kick prescriptions, SN models, mass transfer efficiency and initial mass function. 

%Furthermore, we quantify the impact of observational uncertainties of the cosmic SFR (SFR) and of the metallicity evolution. In particular, we show which metallicity value mostly contributes to the cosmic merger rate density at each given redshift. 

%In order to provide a clear interpretation of the merger rate density evolution with redshift, we create mock simulations. We discussed which is the impact of the merger efficiency and different delay time distributions. We show the feasibility to link properties of the delay time distribution that characterises a binary population to the leading slope of the cosmic merger rate density at $z<1$. 

\section{Methods}
\subsection{Population synthesis}

%{\filippo{what have been the main changes in MOBSE up to now? BH mass distribution? NS mass distribution?}}

We use catalogues of isolated compact binaries from our population-synthesis code {\sc mobse}\footnote{\url{https://mobse-webpage.netlify.app/}}  \citep{mapelli2017, giacobbo2018, giacobbo2018b}. {\sc mobse} includes an up-to-date model for the mass loss rate of massive hot stars, scaling as $\dot M \propto Z^{\beta}$, where $Z$ is the metallicity and $\beta$ depends on the Eddington ratio, as defined in \cite{giacobbo2018}.
%The catalogues of isolated merging compact binaries have been generated with our population-synthesis code {\sc mobse}\footnote{\url{https://mobse-webpage.netlify.app/}}  \citep{mapelli2017, giacobbo2018, giacobbo2018b}. In {\sc mobse}, the mass loss of massive hot stars is described as $\dot M \propto Z^{\beta}$, where $\beta$ is defined as in \cite{giacobbo2018}:
%  \begin{equation}
%    \beta=
%    \begin{cases}
%      0.85, & \text{if}\ \Gamma_{e} < 2/3 \\
%      2.45-2.4\Gamma_{e}, & \text{if}\  2/3\leq{}\Gamma_{e} < 1\\
%      0.05, & \text{if}\ \Gamma_{e} \geq{} 1
%   \end{cases}
%\end{equation}
%where $\Gamma_{e}$ is the electron-scattering Eddington ratio. 

The prescriptions for core-collapse SNe adopted in {\sc mobse} come from \cite{fryer2012} and have been slightly modified to enforce a minimum neutron star (NS) mass of $\approx{}1.23$ M$_\odot$ \citep{giacobbo2020}. %{\nicola I would add 'about' before the number, because the exact value of the mass should be something close to 1.235)} (it was 1.1 M$_\odot$ in the original fitting formulas by \citealt{fryer2012} {\nicola I would remove the last sentence (the one inside the brackets) because it refers only to the rapid model. With the delayed model the minimum mass was already $\sim 1.27$ M$_\odot$)} {\micmap{MM: appendix to be added with new {\sc mobse} features. Nicola's task}}.
Here, we consider both the rapid and the delayed SN model described by \cite{fryer2012}. The two models differ only by the time when the explosion is launched, which is $\lesssim$ 250 ms ($\gtrsim{}500$ ms) after the bounce in the rapid (delayed) model.  According to these models, stars with final carbon-oxygen mass $m_{\rm CO}\gtrsim{}11$ M$_\odot$ collapse to a BH directly. In terms of compact remnant masses, the main difference between the rapid and the delayed model is that the former enforces a mass gap between 2 and 5 M$_\odot$, while the latter does not. 

Following \cite{timmes1996} and \cite{zevin2020}, we compute neutrino mass loss for both NSs and BHs  as
\begin{equation}
    m_\nu{}=\min\left[\frac{\left(\sqrt{1+0.3\,{}m_{\rm bar}}-1\right)}{0.15},\,{}0.5\,{}{\rm M}_\odot\right],
\end{equation}  
where  $m_{\rm bar}$ is the baryonic mass of the compact object. %In addition, we impose that $M_\nu{}$ does not exceed $0.5$ M$_\odot$. Thus,
      The resulting gravitational mass of the compact object is $m_{\rm rem} = m_{\rm bar} - m_\nu{}$.
Prescriptions for pair instability SNe and pulsational pair instability SNe are also implemented, as described in \cite{mapelli2020}. Our treatment for electron-capture SNe is described in \cite{giacobbo2019}.

%{\filippo{I need a recap on natal kicks prescriptions. I mean which is more physical? Giacobbo model only? }}

We consider different SN kick prescriptions, in order to assess their impact on the cosmic merger rate density. %In particular, we implemented two different cases. 
As our fiducial model, we adopt the natal kick prescription proposed by \cite{giacobbo2020}:
%In this model, the natal kick of a compact object (either BH or NS) is expressed as
%\begin{equation}
%v_{\rm kick}=f_{\rm H05}\,{}\left(\frac{m_{\rm ej}}{\langle{}m_{\rm ej,\,{}NS}\rangle{}}\right)\,{}\left(\frac{\langle{}m_{\rm NS}\rangle{}}{m_{\rm rem}}\right),
%\end{equation}
%where $m_{\rm ej}$ is the mass of the ejecta, $\langle{}m_{\rm ej,\,{}NS}\rangle{}$ is the average ejecta mass for a core-collapse SN that leads to the formation of a NS from a single star, $\langle{}m_{\rm NS}\rangle{}$ is the average z in our simulations and $m_{\rm rem}$ is the mass of the specific compact object for which we want to evaluate the natal kick. Finally, $f_{\rm H05}$ is a random number drawn from a Maxwellian distribution with 1-dimensional root mean square $\sigma{}=265$ km s$^{-1}$, according to the fit by \cite{hobbs2005} to the proper motions of young Galactic pulsars.
%They assume that the Maxwellian distribution derived by \cite{hobbs2005} is a good description of NS kicks from single star evolution. They included in their prescription the mass ejecta $m_{\text{ej}}$, since it is reasonable to assume that the magnitude of the kick depends on the total mass ejected during the SN explosion. In order to satisfy linear momentum conservation, they also include a term depending on the mass of the compact object $m_{\text{rem}}$. Hence, the new improved prescription for SN kicks can be expressed as the following equation:
\begin{equation}
\label{eq:kicks}
    v_{\text{kick}} = f_{\text{H05}}\frac{m_{\text{ej}}}{\langle m_{\text{ej}}\rangle}\frac{\langle m_{\text{NS}}\rangle}{m_{\text{rem}}},
\end{equation}
where $f_{\text{H05}}$ is a random value extracted from a Maxwellian distribution with one-dimensional root mean square $\sigma_{\rm 1D} = 265~\si{km s^{-1}}$ \citep{hobbs2005}, $m_{\rm ej}$ is the mass of the ejecta, $m_{\rm rem}$ is the mass of the compact remnant, $\langle m_{\text{NS}}\rangle$ is the average NS mass and $\langle m_{\text{ej}} \rangle$ is the average mass of the ejecta associated with the formation of a NS of  mass $\langle m_{\text{NS}}\rangle$ from single stellar evolution. Equation~\ref{eq:kicks} provides the natal kick for both NSs and BHs, and for both electron-capture and core-collapse SNe. Since BHs that form from direct collapse have $m_{\rm ej}=0$, they receive no kick.  %apart from the one due to neutrino mass loss 
This kick prescription matches the proper motions of young Galactic pulsars \citep{hobbs2005,bray2016,bray2018} and at the same time the merger rate density inferred from LVC \citep{tang2020,giacobbo2020}.

We also consider a simplified model in which the natal kick velocity is randomly drawn from a Maxwellian distribution with fixed one-dimensional root mean square $\sigma_{\rm 1D}$. %, independently on the SN mechanism, i.e. core-collapse SN and electron-capture SN. 
We consider three different values of $\sigma_{\rm 1D}=265$, 150 and 50~km~s$^{-1}$. %{\filippo{why do we select exactly these three values? they represent something in particular?}}.  
In this simple model, the natal kicks of BHs and NSs are drawn from the same Maxwellian distribution, without accounting for direct collapse or fallback. %In this case, we neglect the effect of the fallback, which normally quenches SN kicks for BBHs, especially those which form after direct collapse. {\filippo{citation?}}
%They have implemented the possibility to draw the natal kicks from two Maxwellian distributions with a different root-mean-square: $\sigma_{\rm CCSN}$ and $\sigma_{\rm ECSN}$, for core-collapse and electron-capture SNe, respectively. In the above-mentioned paper, they decided to simulate two extreme cases for the kick of iron core-collapse SNe. In the first case, they draw core-collapse SN kicks from a Maxwellian distribution with $\sigma_{\rm CCSN} = 265$km s$^{-1}$, as derived by \cite{hobbs2005}. In the second case, they draw core-collapse SN kicks from a Maxwellian distribution with $\sigma_{\rm ECSN}=15$km s$^{-1}$, i.e. the same as for electron-capture SNe. 
Furthermore, we consider two alternative kick models. The model F12 (from \citealt{fryer2012}) draws the natal kicks from a Maxwellian distribution with  $\sigma_{\rm {1D}}=265$ km s$^{-1}$  and then modulates the kick magnitude as
\begin{equation}
\label{eq:fallback}
v_{\rm kick}=(1-f_{\rm fb})\,{}f_{\rm H05},
\end{equation}
where $f_{\rm{fb}}$ is the fraction of fallback defined as in \cite{fryer2012}. Finally, the model VG18 (from \citealt{vignagomez2018}) draws the kicks from two different Maxwellian distributions with $\sigma_{\rm {1D}}=265$ and 30 km s$^{-1}$ for NSs born via core-collapse and electron-capture SNe, respectively. Also in this model, the kick is then modulated by the amount of fallback using equation~\ref{eq:fallback}.

In the default version of {\sc mobse}, mass transfer via Roche lobe overflow is described as in \cite{hurley2002}. This yields a nearly conservative mass transfer if the accretor is a non-degenerate star. Here, we introduce also an alternative model in which 
% An important process that concerns binary evolution is Roche lobe overflow (RLOF) and it represents still a source of uncertainty. We explored the impact of two different mass transfer mechanisms during RLOF on the cosmic merger rate density. In the default version {\sc mobse} implements a conservative mass transfer, as described in \cite{hurley2002}. 
the mass accretion rate ($\dot{m}_a$) is described as
%We modified this model by introducing a parameter that describes the mass transfer efficiency $f_{{\rm MT}}$. In this modified version, the accretion rate $\dot{m}_2$ is given by the following equation: 
  \begin{equation}\label{eq:MT}
    \dot{m}_a=\left\{
    \begin{array}{ll}
      f_{\rm MT}\,{}|\dot{m}_d| & \textrm{if the accretor is non-degenerate}\\ \\
      \min{(f_{\rm MT}\,{}|\dot{m}_d|,\dot{m}_{\rm Edd})} & \text{otherwise},
    \end{array}
    \right.
  \end{equation}
where $\dot{m}_d$ is the mass loss rate by the donor star, $\dot{m}_{\rm Edd}$ is the Eddington accretion rate and $f_{\rm MT}\in{}[0,\,{}1]$ is the accretion efficiency. Here, we explore  $f_{\rm MT}=0.1,\,{}0.5$ and 1.

Other binary evolution processes %such as wind mass transfer, tidal evolution, common envelope (CE) and GW energy loss 
are implemented as described in \cite{hurley2002} and \cite{santoliquido2020}. In this work, we assume that the common envelope (CE) ejection efficiency parameter, $\alpha_{\rm CE}$, can assume values from 0.5 and 10, while $\lambda_{\rm CE}$ is derived as described in \cite{claeys2014}.

%$6\times{}10^7$ isolated binaries with $10^7$ systems per each low metallicity values ($Z = 0.0002$, 0.0004, 0.0008, 0.0012, 0.0016, 0.002).  
%For $Z>0.002$ we simulated twice binary systems in order to achieve better statistics. In fact, in previous works ({\filippo{citations}}) has been found that metal-rich stars tend to merge less. Thus for $Z = 0.004$, 0.006, 0.008, 0.012, 0.016, 0.02 we have simulated $1.2\times{}10^8$ isolated binaries with $2\times10^7$ per each metallicity.

In the fiducial model, the mass of the primary star in each binary system is randomly drawn from a \cite{kroupa2001} initial mass function (IMF), with minimum mass $5$~M$_\odot$ and maximum mass $150$~M$_\odot$. For stars with mass $>0.5$~M$_\odot$, the Kroupa IMF behaves as a power law $dN/dm\propto{}m^{-\alpha_{\rm IMF}}$ with $\alpha_{\rm IMF}=2.3$. We also explored %the impact on the cosmic merger rate density of 
different IMF slopes for stars with mass $>0.5$ M$_\odot$. 
%In the default version, {\sc{mobse}} adopts a mass distribution which is given by a power law $m^{-\alpha_{\rm Kroupa}}$, where $\alpha_{{\rm Kroupa}} = 2.3$ for $m>1$ M$_\odot$. We assigned to the same parameter a lower value 
In particular, we consider two cases in which 
$\alpha_{{\rm IMF}} = 2.0$ and 2.7.

Table \ref{tab:models} provides a summary of the different runs performed in this work. 
We have considered 12 different stellar metallicities for each run: $Z = 0.0002$, 0.0004, 0.0008, 0.0012, 0.0016, 0.002, 0.004, 0.006, 0.008, 0.012, 0.016, 0.02. For each run, we have simulated $10^7$ binaries per each metallicity comprised between $Z = 0.0002$ and 0.002, and $2\times{}10^7$ binaries per each metallicity $Z\ge{}0.004$, since higher metallicities are associated with lower BBH and BHNS merger efficiency (e.g. \citealt{giacobbo2018b,klencki2018}). Thus, we have simulated $1.8\times{}10^8$ binaries per each run shown in Table~\ref{tab:models}.

In all runs, the orbital periods, eccentricities and mass ratios of binaries are drawn from \cite{sana2012}. In particular, we derive the mass ratio $q=m_2/m_1$ as $\mathcal{F}(q) \propto q^{-0.1}$ with $q\in [0.1-1]$, the orbital period $P$ from $\mathcal{F}(\Pi) \propto \Pi^{-0.55}$ with $\Pi = \log{(P/\text{day})} \in [0.15 - 5.5]$ and the eccentricity $e$ from $\mathcal{F}(e) \propto e^{-0.42}~~\text{with}~~ 0\leq e \leq 0.9$.

\begin{table}
	\begin{center}
	\caption{Summary of the models.}
	\label{tab:models}
	\begin{tabular}{l c >{\raggedright}p{1.98cm} c c c} % four columns, alignment for each
		\toprule
		Model Name & $\alpha_{\rm CE}$ & Kick Model & SN Model & $f_{\rm MT}$ & $\alpha_{\rm IMF}$ %\vspace{0.1cm}
		\\
		\midrule
		%Figure \ref{fig:mrd} & & & & &\\
		$\alpha0.5$ & 0.5  & Eq.~\ref{eq:kicks} & Delayed & H02 & 2.3 \\		
		$\alpha1$   & 1    & Eq.~\ref{eq:kicks} & Delayed & H02 & 2.3 \\		
		$\alpha2$   & 2    & Eq.~\ref{eq:kicks} & Delayed & H02 & 2.3 \\
        $\alpha3$   & 3    & Eq.~\ref{eq:kicks} & Delayed & H02 & 2.3 \\
        $\alpha5$   & 5    & Eq.~\ref{eq:kicks} & Delayed & H02 & 2.3 \\
        $\alpha7$   & 7    & Eq.~\ref{eq:kicks} & Delayed & H02 & 2.3 \\
        $\alpha10$  & 10   & Eq.~\ref{eq:kicks} & Delayed & H02 & 2.3 
        %\vspace{0.3cm}
        \\\midrule 
        $\alpha1$s265 & 1  & $\sigma_{\rm 1D}=265$ km/s & Delayed & H02 & 2.3\\
        $\alpha5$s265 & 5  & $\sigma_{\rm 1D}=265$ km/s & Delayed & H02 & 2.3 \\
        $\alpha1$s150 & 1  & $\sigma_{\rm 1D}=150$ km/s  & Delayed & H02 & 2.3\\
        $\alpha5$s150 & 5  & $\sigma_{\rm 1D}=150$ km/s  & Delayed & H02 & 2.3 \\
        $\alpha1$s50 & 1  & $\sigma_{\rm 1D}=50$ km/s  & Delayed & H02 & 2.3\\
        $\alpha5$s50 & 5  & $\sigma_{\rm 1D}=50$ km/s  & Delayed & H02 & 2.3 \\
        $\alpha$1F12 & 1 & Eq. \ref{eq:fallback}& Delayed & H02 & 2.3 \\
        $\alpha$5F12 & 5 & Eq. \ref{eq:fallback} & Delayed & H02 & 2.3 \\
        $\alpha$1VG18 & 1 & $\sigma_{\rm{high}} = 265$ km/s $\sigma_{\rm{low}} = 30$ km/s & Delayed & H02 & 2.3 \\
        $\alpha$5VG18 & 5 & $\sigma_{\rm{high}} = 265$ km/s $\sigma_{\rm{low}} = 30$ km/s & Delayed & H02 & 2.3
        %\vspace{0.3cm}
        \\\midrule
        $\alpha1$R  & 1 & Eq.~\ref{eq:kicks} & Rapid & H02 & 2.3 \\		
        $\alpha5$R  & 5 & Eq.~\ref{eq:kicks} & Rapid & H02 & 2.3 
        %\vspace{0.3cm}
        \\\midrule
        $\alpha1$MT0.1 & 1 & Eq.~\ref{eq:kicks} & Delayed & 0.1 & 2.3\\
        $\alpha1$MT0.5 & 1 & Eq.~\ref{eq:kicks} & Delayed & 0.5 & 2.3\\
        $\alpha1$MT1.0 & 1 & Eq.~\ref{eq:kicks} & Delayed & 1.0 & 2.3\\
        $\alpha5$MT0.1 & 5 & Eq.~\ref{eq:kicks} & Delayed & 0.1 & 2.3\\
        $\alpha5$MT0.5 & 5 & Eq.~\ref{eq:kicks} & Delayed & 0.5 & 2.3\\
        $\alpha5$MT1.0 & 5 & Eq.~\ref{eq:kicks} & Delayed & 1.0 & 2.3\\
        $\alpha10$MT0.1 & 10 & Eq.~\ref{eq:kicks} & Delayed & 0.1 & 2.3\\
        $\alpha10$MT0.5 & 10 & Eq.~\ref{eq:kicks} & Delayed & 0.5 & 2.3\\
        $\alpha10$MT1.0 & 10 & Eq.~\ref{eq:kicks} & Delayed & 1.0 & 2.3%\vspace{0.3cm}
        \\\midrule
        $\alpha1$IMF2.0  & 1 & Eq.~\ref{eq:kicks} & Delayed & H02 & 2.0\\
        $\alpha1$IMF2.7  & 1 & Eq.~\ref{eq:kicks} & Delayed & H02 & 2.7\\
        $\alpha5$IMF2.0  & 5 & Eq.~\ref{eq:kicks} & Delayed & H02 & 2.0\\
        $\alpha5$IMF2.7  & 5 & Eq.~\ref{eq:kicks} & Delayed & H02 & 2.7
       %\vspace{0.1cm}
       \\
		\bottomrule
	\end{tabular}
	\end{center}
%	\shiftleft{
	\footnotesize{Column~1: model name. Column~2: parameter $\alpha_{\rm CE}$ of the CE. Column~3: kick model; runs $\alpha1$s265/$\alpha5$s265, $\alpha1$s150/$\alpha5$s150 and $\alpha1$s50/$\alpha5$s50  have natal kicks drawn from a Maxwellian distribution with root mean square $\sigma_{\rm 1D} = 265$, 150 and 50 km s$^{-1}$, respectively;  runs $\alpha$1F12 and $\alpha{}$5F12 adopt the natal kick model in eq.~\ref{eq:fallback}; runs $\alpha{}1$VG18 and $\alpha{}5$VG18 assume the same model as \cite{vignagomez2018}; in all the other models, the kicks are calculated as in eq.~\ref{eq:kicks}. Column~4: core collapse SN model; models $\alpha1$R and $\alpha5$R adopt the rapid model from \cite{fryer2012}, while all the other models adopt the delayed model from the same authors. Column~5: accretion efficiency $f_{\rm MT}$ onto a non-degenerate accretor; H02 means that we follow the same formalism as in \cite{hurley2002}. For the other models, see eq.~\ref{eq:MT}. Column~6: slope of the IMF; models $\alpha_{\rm IMF}$ of the IMF  for $m>0.5$ M$_\odot$; $\alpha1$K2.0, $\alpha5$K2.0 ($\alpha1$K2.7, $\alpha5$K2.7)  have  $\alpha_{\rm IMF}=2.0$ ($\alpha_{\rm IMF}=2.7$). All the other models assume the "standard" slope $\alpha_{\rm IMF}=2.3$ \citep{kroupa2001}.}%}
\end{table}

\subsection{Cosmic merger rate density}
\label{sec:mrd}
%We already presented the methodology we employed to evaluate the cosmic merger rate density of compact binaries in \cite{santoliquido2020}. Here we provide some more details. 

We model the cosmic merger rate density $\mathcal{R}(z)$ following \cite{santoliquido2020}:
\begin{equation}
\label{eq:mrd}
   \mathcal{R}(z) = \frac{{\rm d}~~~~~}{{\rm d}t_{\rm lb}(z)}\left[\int_{z_{\rm max}}^{z}\psi(z')\,{}\frac{{\rm d}t_{\rm lb}(z')}{{\rm d}z'}\,{}{\rm d}z'\int_{Z_{\rm min}}^{Z_{\rm max}}\eta(Z) \,{}\mathcal{F}(z',z, Z)\,{}{\rm d}Z\right],
\end{equation}
where $t_{\rm lb}(z)$ is the look-back time at redshift $z$, $Z_{\rm min}$ and $Z_{\rm max}$ are the minimum and maximum metallicity, $\psi{}(z')$ is the cosmic SFR density at redshift $z'$, $\mathcal{F}(z',z,Z)$ is the fraction  of compact binaries that form at redshift $z'$ from stars with metallicity $Z$ and merge at redshift $z$, and $\eta(Z)$ is the merger efficiency, namely the ratio between the total number $\mathcal{N}_{\text{TOT}}(Z)$ of compact binaries (formed from a coeval population) that merge within an Hubble time ($t_{{\rm H}_0} \lesssim 14$ Gyr) and the total initial mass $M_\ast{}(Z)$ of the simulation with metallicity $Z$:
\begin{equation}
\label{eq:eta}
    \eta (Z) = f_{\rm bin}f_{\rm IMF}  \frac{\mathcal{N}_{\text{TOT}}(Z)}{M_\ast{}(Z)},
\end{equation}
where $f_{\rm bin} = 0.5$ is the binary fraction, and $f_{\rm IMF}$ is a correction factor that takes into account that only stars with mass $m>5$ M$_\odot$ are simulated. This parameter depends on the adopted IMF, in particular $f_{\rm IMF} = 0.483$, 0.285  and 0.123 when $\alpha_{\rm IMF} = 2.0$, 2.3 and 2.7 respectively.  The cosmological parameters used in equation~\ref{eq:mrd} are taken  from \cite{planck2016}. The maximum considered redshift in equation~\ref{eq:mrd} is $z_{\rm max}=15$.

%In first approximation, the cosmic MRD is driven by the SFR density. 
The SFR density $\psi(z)$ is  described as \citep{madau2017}:
\begin{equation}
\label{eq:sfrd}
 \psi(z) = 0.01\,{}\frac{(1+z)^{2.6}}{1+[(1+z)/3.2]^{6.2}}~\text{M}_\odot\,{}\text{Mpc}^{-3}\,{}\text{yr}^{-1}.
\end{equation}
As detailed in \cite{santoliquido2020}, we assume that the errors follow a log-normal distribution with mean $\log{\psi{}(0)}=-2$ and standard deviation $\sigma_{\log{\psi}}=0.2$. %, taking into account the typical $1\,{}\sigma{}$ error bars on single data points (see Figure~9 of \citealt{madaudickinson2014}). 

The normalisation of equation~\ref{eq:sfrd} is obtained for a Kroupa IMF with $\alpha_{\rm IMF}=2.3$. When we vary the  slope of the IMF, % and then we evaluate the cosmic merger rate density, 
we have to change the normalisation of eq.~\ref{eq:sfrd} \citep{madaudickinson2014}.  Thus, we re-scale the normalisation by multiplying equation~\ref{eq:sfrd} by a factor 0.58 and 2.40 for $\alpha_{\rm IMF}=2.0$ and 2.7, respectively \citep[see, e.g.,][]{klencki2018}.

 The average stellar metallicity $\mu{}(z)$ evolves with redshift as
%We describe the evolution of the average stellar metallicity as a function of redshift $\mu{}(z)$, as follows.
\begin{equation}
\label{eq:metmod}
 \mu(z) =  \log{ \left(\frac{Z(z)}{Z_\odot}\right)} = \log{(a)}\,{}+\,{}b\,{}z,
\end{equation}
where $a=1.04\pm{}\,{0.14}$ and $b=-0.24\pm{}0.14$,  based on observational results \citep{gallazzi2008,decia2018}. We refer to \cite{santoliquido2020} for a discussion on the choice of $a$ and $b$. %In the above equation, the slope $b$ comes from \cite{decia2018}, who provide a fit to the metallicity evolution of a large sample of damped Lyman$-\alpha$ (DLA) systems with redshift between 0 and 5. The original fit by \cite{decia2018} yields a metallicity $Z(z = 0)=0.66$ Z$_\odot$, which is low compared to the average stellar  metallicity measured at redshift zero (see for instance the discussion in \citealt{madaudickinson2014}). Hence, in equation~\ref{eq:metmod}, we have re-scaled the fitting formula provided by \cite{decia2018} to yield $Z(z = 0) = (1.04\pm{}\,{0.14})~{\rm Z}_\odot$,  where $Z_\odot=0.019$, consistent with the average metallicity of  galaxies at $z\sim{}0$ from the Sloan Digital Sky Survey \citep{gallazzi2008}. The quoted uncertainties on both $a$ and $b$ are at 1 $\sigma{}$. We assume that the observational values follow a Gaussian distribution, as it has been done in the original papers by \cite{gallazzi2008} and \cite{decia2018}. 

We model the distribution of stellar metallicities $\log{(Z/{\rm Z}_\odot)}$ at a given redshift as a normal distribution with mean value $\mu{}(z)$ from equation~\ref{eq:metmod} and standard deviation $\sigma_{Z} = 0.20$ dex as our fiducial value: 
\begin{equation}
\label{eq:pdf}
p(z', Z) = \frac{1}{\sqrt{2 \pi\,{}\sigma_{Z}^2}}\,{} \exp\left\{{-\,{} \frac{\left[\log{(Z/{\rm Z}_\odot)} - \mu(z')\right]^2}{2\,{}\sigma_{Z}^2}}\right\}.
\end{equation}
%We assume $\sigma{}_{Z}=0.20$ dex, based on the metallicity spread found in cosmological simulations (e.g., {\sc eagle}, \citealt{artale2019}). 
 In Section \ref{sec:metandsfr}, we discuss the impact of a different choice of $\sigma_Z$ on the merger rate density. Previous works have calculated the metallicity evolution based on a number of different assumptions and have shown its importance for the estimate of the merger rate (e.g., \citealt{dominik2013,dominik2015,belczynski2016,lamberts2016,mapelli2017,mapelli2018,neijssel2019,baibhav2019,chruslinska2019,chruslinska2020}). %(see also \citealt{chruslinska2019,chruslinska2020}). 

The fraction  of compact binaries that form at redshift $z'$ from stars with metallicity $Z$ and merge at redshift $z$ is thus given by
\begin{equation}
\label{eq:fraction}
\mathcal{F}(z',z,Z)=\frac{\mathcal{N}(z',z,Z)}{\mathcal{N}_{\text{TOT}}(Z)}\,{}p(z', Z),
\end{equation}
where $\mathcal{N}(z',z,Z)$ is the total number of compact binaries that merge at redshift $z$ and form from stars with metallicity $Z$ at redshift $z'$.

%Another important task of this work is to determine the impact on the cosmic merger rate density of observational uncertainties. 
We performed $2\times 10^{3}$ realisations of equation~\ref{eq:mrd} per each model in Table~\ref{tab:models}. %For $10^3$ realisations, 
In each realisation, we randomly draw the normalisation value of the SFR density (equation~\ref{eq:sfrd}), and %for the remaining $10^3$ realisations we draw 
the intercept and the slope of the average metallicity (equation~\ref{eq:metmod}) from three  Gaussian distributions with mean (standard deviation) equal to $\log{\psi{}(0)}=-2$ ($\sigma_{\log{\psi}}=0.2$), $a=1.04$ ($\sigma_a=0.14$) and $b=-0.24$ ($\sigma_b=0.14$), respectively. For simplicity, the value of the intercept and that of the slope are drawn separately, assuming no correlation. %We thus kept separated the impact of SFR density normalisation and metallicity evolution. %These results are shown in Figure \ref{fig:err}.

%We integrate the quantity given in equation~\ref{eq:fraction} over all possible metallicities and we multiply it by the SFR density at redshift $z'$, integrating again over all redshift. This last operations eventually provide the cosmic merger rate density:
%\begin{equation}
%\label{eq:mrd}
%   \mathcal{R}(z) = \frac{d~~~~~}{dt_{\rm lb}(z)}\left[\int_{z_{\rm max}}^{z}\psi(z')\,{}\frac{{\rm d}t_{lb}(z')}{{\rm d}z'}\,{}{\rm d}z'\int_{Z_{\rm min}}^{Z_{\rm max}}\mathcal{F}(z',z, Z)\,{}{\rm d}Z\right]
%\end{equation}
%where $t_{\rm lb}(z)$ is the look-back time at redshift $z$, $Z_{\rm min}$ and $Z_{\rm max}$ are the minimum and maximum metallicities, $\eta{}(Z)$ is the merger efficiency at metallicity $Z$. To calculate the lookback time we take the cosmological parameters ($H_{0}$, $\Omega_{\rm M}$ and $\Omega_{\Lambda}$)  from \cite{planck2016}. The maximum considered redshift in equation~\ref{eq:mrd} is $z_{\rm max}=15$, which we assume to be the epoch of formation of the first stars. 

%This model is implemented in the new python script {\sc cosmo}$\mathcal{R}${\sc ate}, which allows us to calculate up to $10^{3}$ models per day on a single core. 

%%%%%%%%%%%%%%%%%%%%%%%%%%%%%%%%%%%%%%%%%%%%%%%%%%%%%

\section{Results}
\subsection{Merger efficiency}
%%%%%%%%%%%%%%%%%%%%%%%%%%%%%FIGURE%%%%%%%%%%%%%%%%%%%%%%%%%%%%%%%%%%%%%%%%%%%%%%%
\begin{figure}
	\includegraphics[width= \columnwidth]{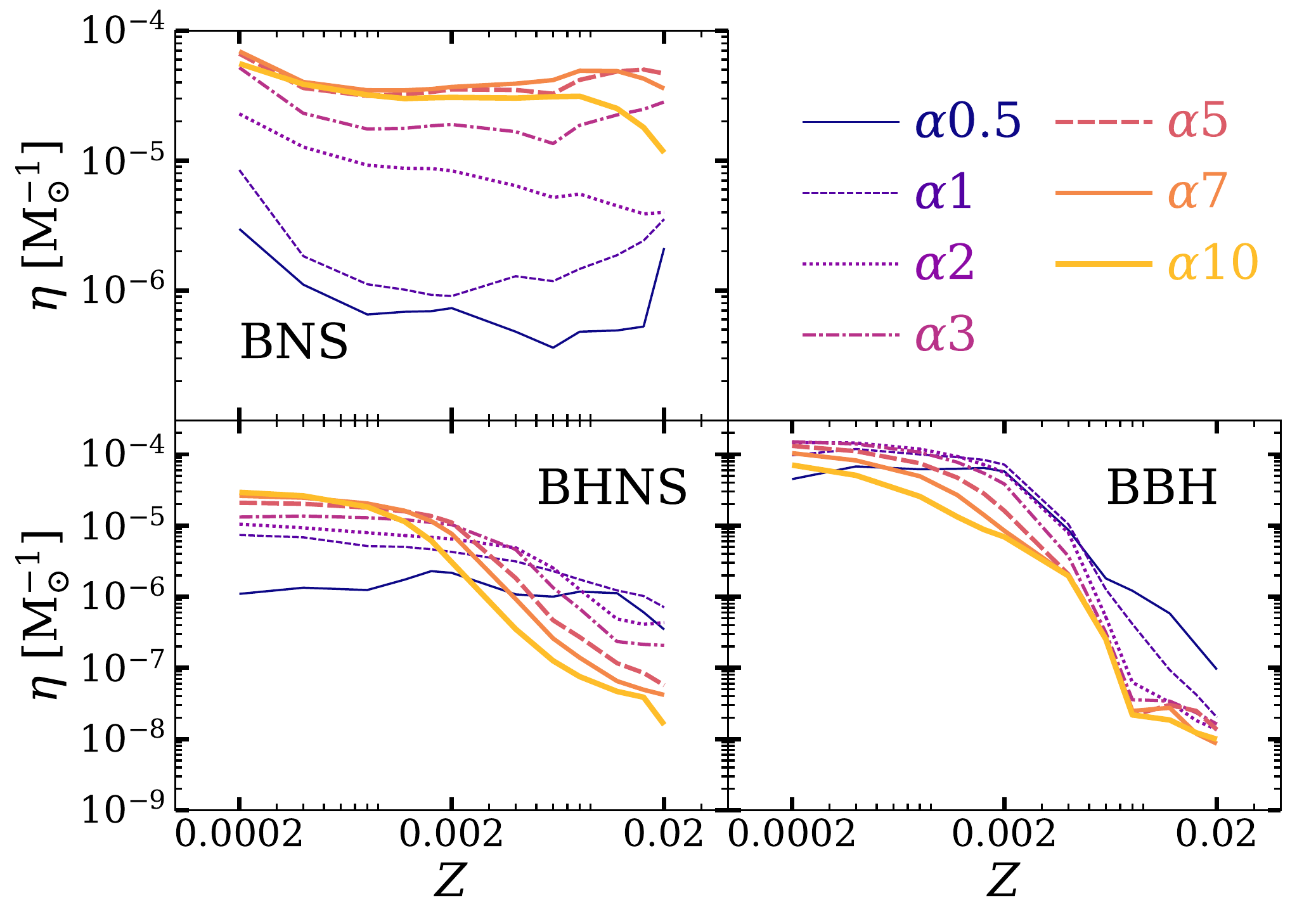}
    \caption{Merger efficiency $\eta$ as a function of progenitor’s metallicity for models $\alpha0.5$ to $\alpha10$. See Table~\ref{tab:models} for further details.
    }
    \label{fig:eta}
\end{figure}
%%%%%%%%%%%%%%%%%%%%%%%%%%%%%%%%%%%%%%%%%%%%%%%%%%%%%%%%%%%%%%%%%%%%%%%%%%%%%%%%%%%%

Figure \ref{fig:eta}  shows the merger efficiency $\eta(Z)$ as defined in equation~\ref{eq:eta}, as a function of progenitor's metallicity and for different values of the $\alpha_{\rm CE}$ parameter.  The BNS merger efficiency (hereafter, $\eta_{\rm BNS}$) mildly depends on the metallicity of the progenitor star, as already found by previous works \citep[e.g.,][]{mapelli2010,Chakrabarti2017,giacobbo2018,klencki2018,chruslinska2018, neijssel2019}. The behaviour of $\eta{}_{\rm BNS}$ as a function of metallicity is different for different values of $\alpha_{\rm CE}$. For example, for $\alpha_{\rm CE}{}=1$, $\eta{}_{\rm BNS}$ has a U$-$shaped trend with metallicity and has a minimum at $Z=0.002$, while, for $\alpha_{\rm CE}{}=2$, $\eta{}_{\rm BNS}$ decreases almost monotonically from $Z=0.0002$ to $Z=0.02$.
%Going from a low value of the $\alpha_{\rm CE}$ parameter ($\alpha_{\rm CE} = 0.5$) to greater values, $\eta$ tends to follow a flat trend with $Z$ {\micmap COSA SIGNIFICA FLAT TREND? vuoi forse dire che l'efficiency aumenta senza che la curva cambi slope? mi sembra diverso da quello che hai scritto}. 
The BNS merger efficiency changes by less than one order of magnitude with $Z$, while it increases by two orders of magnitude with increasing $\alpha_{\rm CE}$. %Thus, we can state from this plot that what triggers the mergers of BNSs is not the progenitor metallicity but the efficiency of CE phase in shrinking the binary system. 
Thus, the BNS merger efficiency is strongly affected by the CE parameter $\alpha_{\rm CE}{}$ and only mildly affected by metallicity.

%The BHNS merger  efficiency is characterised by a more prominent dependence on $Z$. For $\alpha_{\rm CE}{}\ge{}5$, $\eta_{\rm BHNS}$ decreases by three order of magnitude going from $Z=0.0002$ up to $Z=0.02$. 

The behaviour of the BHNS merger efficiency (hereafter, $\eta_{\rm BHNS}$) as a function of metallicity dramatically depends on the value of $\alpha_{\rm CE}$. By decreasing the value of $\alpha_{\rm CE}$, $\eta{}_{\rm BHNS}$ progressively decreases at low $Z$ and increases at high $Z$.
%This trend looks like to converge for $\alpha_{\rm CE} \rightarrow 0$. In fact, 
For large values of $\alpha_{\rm CE}$ ($\ge{}5$),  $\eta_{\rm BHNS}$ decreases by three orders of magnitude going from $Z=0.0002$ up to $Z=0.02$, while for $\alpha_{\rm CE}=0.5$ $\eta_{\rm BHNS}$ is almost independent of $Z$.

This can be physically explained by an interplay between stellar winds and CE. A small value of $\alpha_{\rm CE}$ ($\alpha_{\rm CE} \lesssim 1$) means inefficient CE ejection: the binary has to shrink a lot before the envelope is ejected. At low $Z$, inefficient CE ejection suppresses the merger of small BHs (with mass $< 10$ M$_\odot$),  because their progenitor stars retain large envelopes and merge during CE, before giving birth to BHNSs. In contrast, at solar metallicity, stellar winds peel off stars and their envelopes are relatively small, making it difficult for CE to harden the system enough to merge by GW emission. Hence, an inefficient CE ejection tends to boost mergers of low-mass BHs at solar metallicity, by efficiently shrinking their progenitor binaries.
%while at solar metallicity the mergers of these systems are indeed triggered. 
%In other words, at metallicity values close to solar an inefficient CE ejection can shrink and consequently let merge binaries formed by stars which have a small radius, as a consequence of the high mass loss. 
 
%{\micmap QUI ADESSO BISOGNA DARE UN'INTERPRETAZIONE}

%In this case, $\eta$ shows a peculiar and more complicated trend with $\alpha$ CE values. For $Z\lesssim0.002$, the greater is $\alpha$ the greater is $\eta$. This trend is in line with BNS merger efficiency, then it is reverted when we consider $Z \gtrsim 0.002$. For  $\alpha < 1$ this tends to be suppressed. In fact, $\eta$ for $\alpha = 0.5$ shows almost no dependence with $Z$. 
%{\micmap per come lo dici tu sembra che $\alpha=0.5$ sia diverso da tutti gli altri. Invece io vedo un chiarissimo trend: la differenza tra alta e bassa metallicit\`a si risuce costantemente al diminuire di alpha}

The BBH merger efficiency (hereafter, $\eta_{\rm BBH}$) strongly depends on progenitor's metallicity and is only mildly affected by $\alpha_{\rm CE}$. $\eta_{\rm BBH}$  decreases by three--four orders of magnitude from the lowest to the highest considered metallicity. Lower values of $\alpha_{\rm CE}$ result in higher values of $\eta_{\rm BBH}$, with the exception of the case with $\alpha_{\rm CE}=0.5$ and $Z=0.0002$.

%BBH merger efficiency also shows an opposite trend with respect to BNS one. In fact, it decreases with increasing $\alpha_{\rm CE}$.

\subsection{Common envelope} 
\label{sec:ce}

%%%%%%%%%%%%%%%%%%FIGURE%%%%%%%%%%%%%%%%
\begin{figure}
\centering
\includegraphics[width=\columnwidth ]{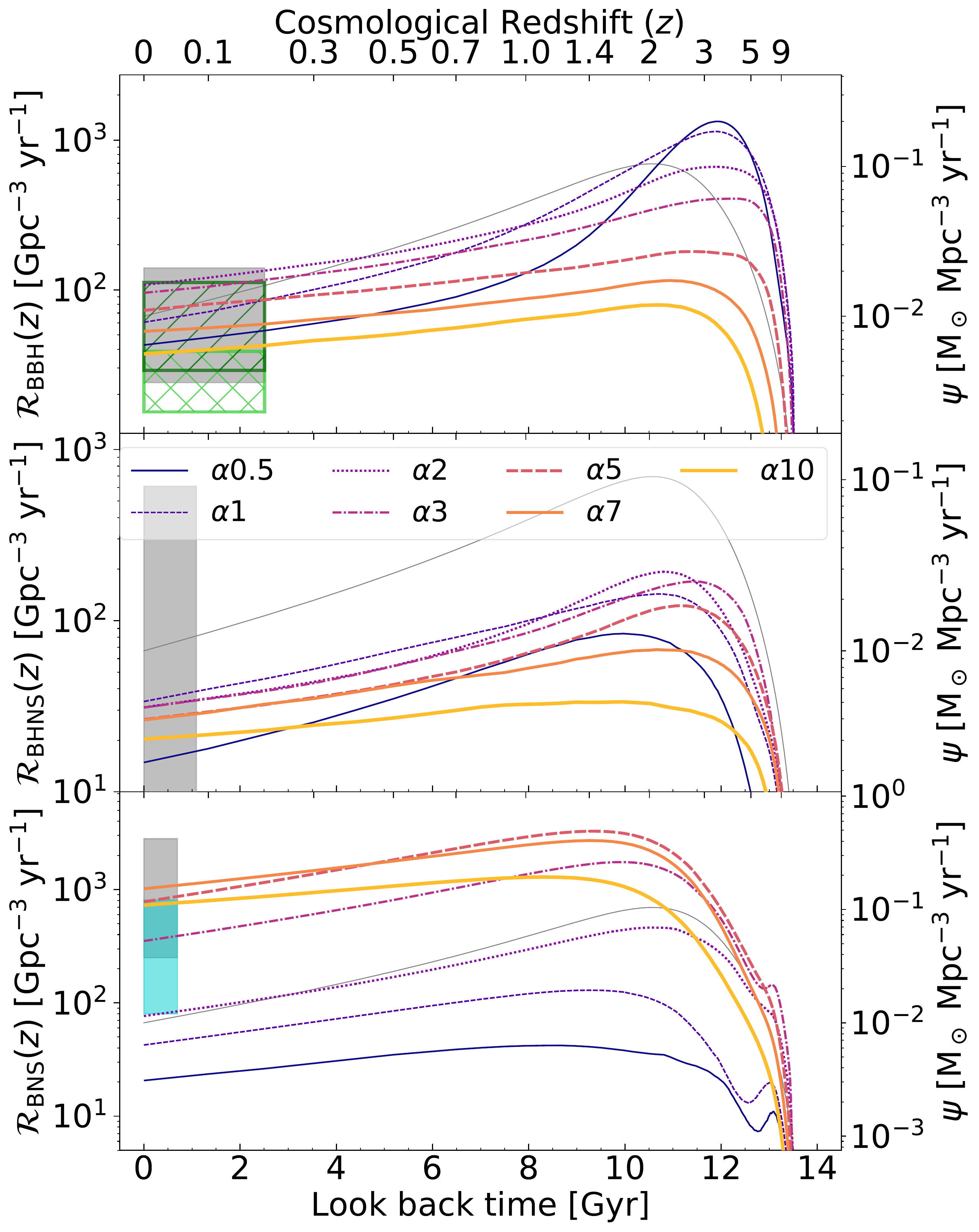} 
\caption{Left $y-$axis: Evolution of the merger rate density $\mathcal{R}(z)$ for BBHs (top), BHNSs (centre) and BNSs (bottom) in the comoving frame, as a function of the look-back time (lower $x-$axis) and of the redshift (upper $x-$axis). We vary $\alpha_{\rm CE}$ from 0.5  (model $\alpha0.5$) to 10 (model $\alpha10$).  For both BBHs and BHNSs, the grey shaded area shows the 90\% credible interval of the local merger rate density, as inferred from the first two observing runs of the LVC \protect\citep{abbottO2,abbottO2popandrate}. For BBHs, we consider the union of the rates obtained with model A, B and C in \protect\cite{abbottO2popandrate}.  For BNSs, the grey shaded area shows the merger rate density estimated in \protect\cite{abbottGW190425}. 
The hatched green areas in the upper panel show  the local BBH  merger rate density inferred including O3a events \protect\citep{abbottpopO3a}. In particular, the dark-green and  light-green hatched areas show the 90\% credible interval calculated including and excluding GW190814-like events, respectively \protect\citep{abbottpopO3a}. Finally, the cyan shaded area in the lower panel shows the 90\% credible interval of the local BNS merger rate density, as estimated by \protect\cite{abbottpopO3a}. The width of the shaded and hatched areas on the $x-$axis  corresponds to the instrumental horizon obtained by assuming BBHs, BHNS, BNSs  of mass $(10+10), (1.4-5)$ and $(1.4-1.4)$ M$_\odot$ respectively and O2 sensitivity \protect\citep{abbott2018observingscenario}. Right $y-$axis and grey solid thin line: SFR density evolution (equation \protect\ref{eq:sfrd}).} 
\label{fig:mrd}
\end{figure}
%%%%%%%%%%%%%%%%%%FIGURE%%%%%%%%%%%%%%%%

Figure \ref{fig:mrd} shows the cosmic merger rate density $\mathcal{R}(z)$ as a function of redshift for the same values of the CE parameter as shown in Figure~\ref{fig:eta}. The BNS merger rate density is up to two orders of magnitude higher for large values of $\alpha_{\rm CE}$ than for low values. This trend can be easily explained by looking at the merger efficiency (Figure~\ref{fig:eta}): for BNSs, larger values of $\alpha_{\rm CE}$ translate into higher merger efficiency.
%The normalization of the BNS merger rate density as a function of redshift can be interpreted based on the merger efficiency $\eta{}_{\rm BNS}$. In fact,  the merger rate density increases significantly for large values of $\alpha_{\rm CE}$, because the merger efficiency grows with $\alpha_{\rm CE}{}$. Hence, the local merger rate density is consistent with the 90\% credible interval inferred by the LIGO-Virgo collaboration only when $\alpha_{\rm CE} >3$ is adopted. 

The top panel of Figure \ref{fig:mrd} shows the merger rate density of
BBHs. In the local Universe,  $\mathcal{R}_{\rm BBH}(z)$ changes by a factor of $2-3$ if we change $\alpha_{\rm CE}$. Thus, the impact of $\alpha_{\rm CE}$ on the local BBH merger rate is smaller than in the case of BNSs. Moreover, models with large $\alpha_{\rm CE}$ result in lower BBH merger rates, with an opposite trend with respect to BNSs. These differences are also explained by  the behaviour of the merger efficiency  at different $\alpha_{\rm CE}$ (Figure~\ref{fig:eta}).

The merger rate density of BHNSs follows an evolution similar to that of BBHs: lower values of $\alpha_{\rm CE}$ give higher merger rates (with the exception of $\alpha_{\rm CE}=0.5$) and the difference between models with different $\alpha_{\rm CE}$ is only a factor of $\sim{}2$ in the local Universe. As for BNSs and BBHs, this trend can be explained by looking at the merger efficiency. From now on, we consider $\alpha_{\rm CE}=1$ and 5 as our fiducial cases.

In Figure \ref{fig:mrd} and following, we show the 90\% credible intervals inferred by the LVC. The grey boxes represent the values inferred from the first and second observing runs (GWTC-1, \citealt{abbottO2}) for BBHs ($\mathcal{R}_{\rm{BBH}} = 24 - 140$ Gpc$^{-3}$ yr$^{-1}$), BHNSs ($\mathcal{R}_{\rm{BHNS}} \leq 610$ Gpc$^{-3}$ yr$^{-1}$)  and BNSs ($\mathcal{R}_{\rm{BNS}} = 250 - 2810$ Gpc$^{-3}$ yr$^{-1}$). 
From GWTC-2 \citep{abbottO3a}, we considered the 90\% credible intervals inferred for BBHs, with and without taking into account GW190814, which are $\mathcal{R}_{\rm{BBH}} = 23.9^{+14.9}_{-8.6}$ Gpc$^{-3}$ yr$^{-1}$ (hatched light-green box) and $\mathcal{R}_{\rm{BBH}} = 58^{+54}_{-29}$ Gpc$^{-3}$ yr$^{-1}$ (hatched dark-green box), respectively. The cyan box is the updated 90\% credible interval inferred for BNSs, which is equal to $\mathcal{R}_{\rm{BNS}} = 320^{+490}_{-240}$ Gpc$^{-3}$ yr$^{-1}$.

%{\filippo{another important concept, $\alpha_{\rm CE}$ on BNS MRD does not change its slope while it does it for BBH...how can I explain this phenomenon and why it happens?} \micmap{WITH METALLICITY!!!! mentre per le NS il trend rimane sempre pi\`u o meno quello perch\'e non hai grossi cambiamenti con la metallicit\`a, per i BBH quando sali in redshift lo slope diventa pi\`u ripido perch\'e entri in un regime dove la star formation avviene prevalentemente a metallicit\ `a pi\`u bassa. prova a scriverlo}} 

\subsection{Natal kicks}

%%%%%%%%%%%%%%%%%%FIGURE%%%%%%%%%%%%%%%%
\begin{figure}
\centering
\includegraphics[width= \columnwidth]{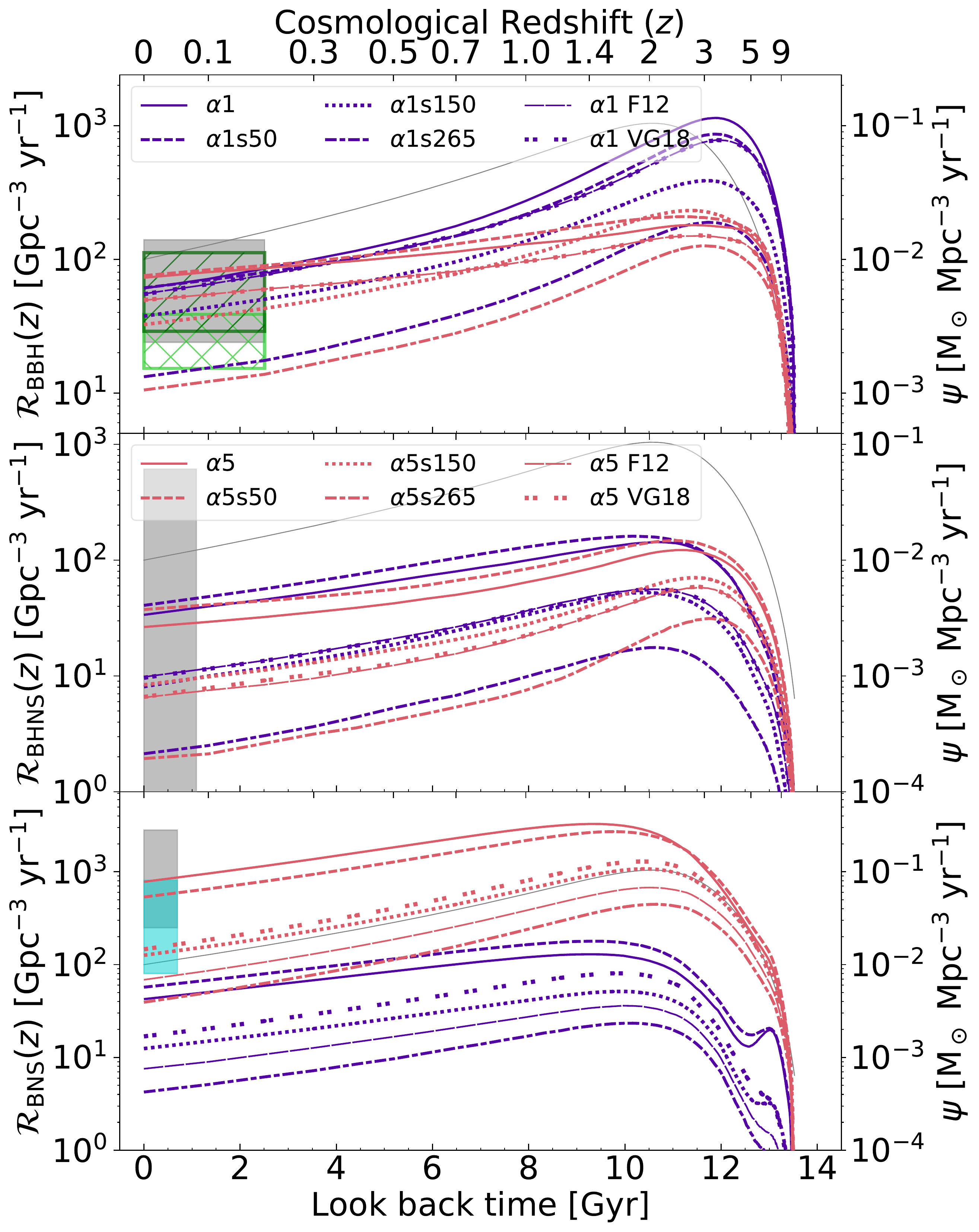}
\caption{Merger rate density of BBHs (top), BHNSs (centre) and BNSs (bottom). Same as Figure~\ref{fig:mrd}, but we compare different natal kicks. %{\celeste Please check that the cSFR (grey lines) in the Figures 2,3,4,5, and 6 seems to reach a different value for the BNS panels at zmax.}
} 
\label{fig:kicks}
\end{figure}
%%%%%%%%%%%%%%%%%%FIGURE%%%%%%%%%%%%%%%%

Figure \ref{fig:kicks} shows that the higher the natal kick is, the lower is the merger rate density at each given redshift, for BBHs, BHNSs and BNSs. In fact, high natal kicks tend to disrupt the binary system.  SN kicks drawn from a Maxwellian distribution with $\sigma_{\rm 1D} = 50~$kms$^{-1}$ yield a merger rate density similar to that given by equation~\ref{eq:kicks}. %{\yann{Maybe explain why the two values alpha1 and alpha5 are the cases selected to compare values of common enveloppe in most of the following sections.}} %This is expected since this SN kick prescription provides an average velocity kick  $\sim50$ kms$^{-1}$, as it has been shown in \cite{giacobbo2019}. 

As expected from the binary binding energy, the effect of different SN natal kick prescriptions is higher for BNSs, where there is a difference up to an order of magnitude if we consider natal kicks drawn from a Maxwellian with $\sigma_{\rm 1D} = 265$km s$^{-1}$ with respect to $\sigma_{\rm 1D} = 50$ km s$^{-1}$.

Only
models %in which %$v_{\rm kick}\propto{}m_{\rm ej}$ (like the one in equation~\ref{eq:kicks}) 
with relative low natal kicks and large values of $\alpha_{\rm CE}$ (like $\alpha{}5$, $\alpha{}5$s50, $\alpha{}5$s150, and $\alpha{}5$VG18) 
%can match the merger rate density inferred from LVC data when considering both O1, O2 and O3a
are inside the 90\% credible interval of GWTC-2 \citep{abbottO3a,abbottpopO3a}. On the other hand, a single Maxwellian curve with $\sigma_{\rm 1D} = 50$ km s$^{-1}$ (e.g., models $\alpha{}1$s50 and $\alpha{}5$s50) is in tension with the observed proper motions of young pulsars in our Galaxy \citep{hobbs2005,verbunt2017,pol2019}. Hence, only models $\alpha{}5$, $\alpha{}5$s150 and $\alpha{}5$VG18 are still consistent with both pulsars' proper motions and GW data.

\subsection{Core-collapse SN model}

%%%%%%%%%%%%%%%%%%FIGURE%%%%%%%%%%%%%%%%
\begin{figure}
\centering
\includegraphics[width= \columnwidth]{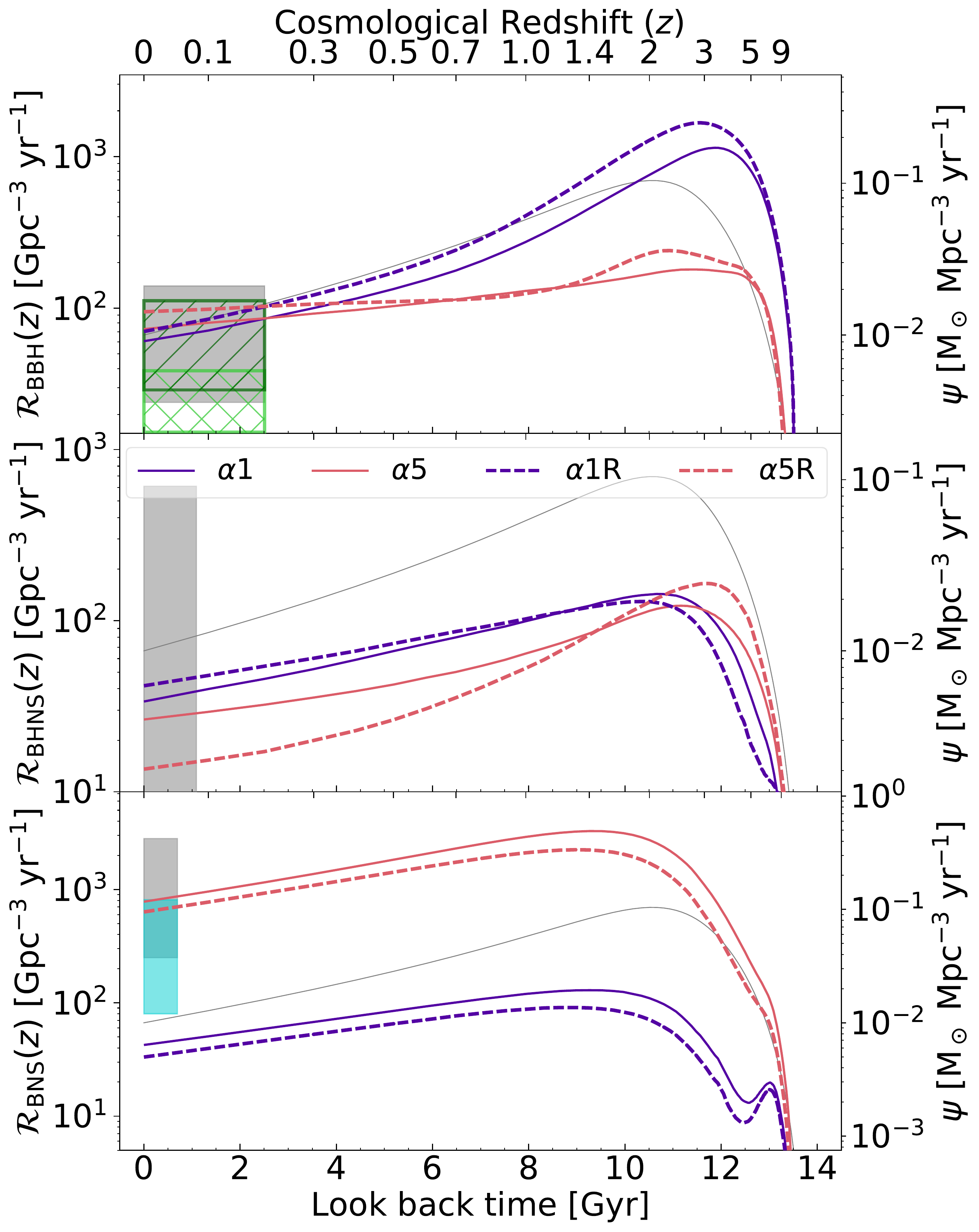}
\caption{Merger rate density of BBHs (top), BHNSs (centre) and BNSs (bottom). Same as Figure~\ref{fig:mrd}, but we compare the rapid and delayed core-collapse SN models.}
\label{fig:snmodel}
\end{figure}
%%%%%%%%%%%%%%%%%%FIGURE%%%%%%%%%%%%%%%%

Choosing the delayed or the rapid core-collapse SN model has a minor impact on the cosmic merger rate density (Figure~\ref{fig:snmodel}). The delayed model slightly enhances $\mathcal{R}_{\rm {BNS}}(z)$, because it produces more massive NSs which can merge on a shorter timescale. For the same reason, the delayed model slightly suppresses $\mathcal{R}_{\rm{BBH}}(z)$, because it produces a number of low-mass BHs ($3-5$ M$_\odot$), which merge on a longer timescale than more massive BHs. For BHNSs, the effect of the core-collapse SN model depends on the choice of the $\alpha_{\rm CE}$ parameter.

\subsection{Mass accretion efficiency}
%%%%%%%%%%%%%%%%%%FIGURE%%%%%%%%%%%%%%%%
\begin{figure}
\centering
\includegraphics[width=\columnwidth]{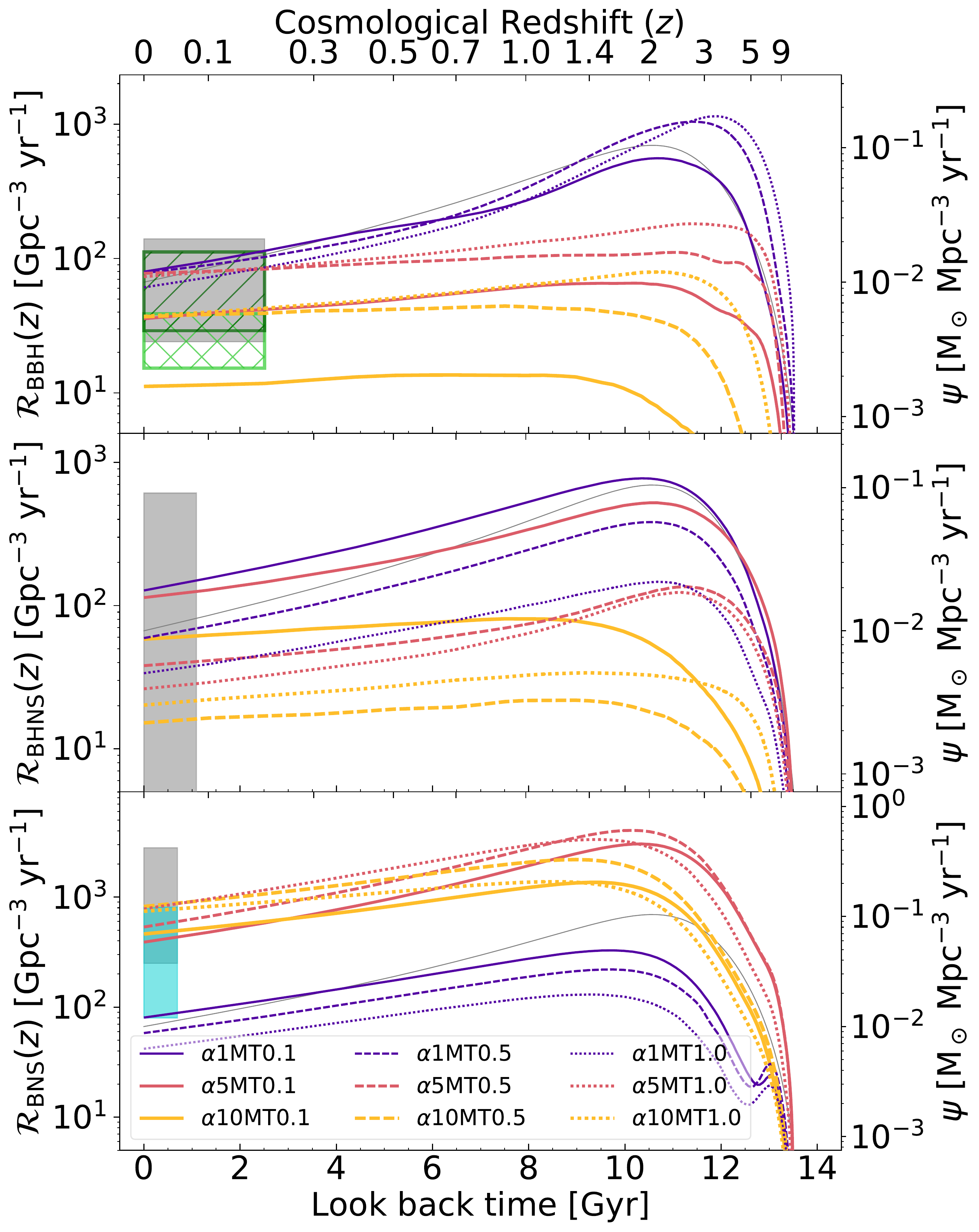}
\caption{Merger rate density of BBHs (top), BHNSs (centre) and BNSs (bottom).  Same as Figure~\ref{fig:mrd}, but we compare different values of the accretion efficiency parameter $f_{\rm MT}$.}
\label{fig:mt}
\end{figure}
%%%%%%%%%%%%%%%%%%FIGURE%%%%%%%%%%%%%%%%

Figure \ref{fig:mt} shows the impact of different values of the mass accretion efficiency  on the cosmic merger rate density.  Lower values of $f_{\rm MT}$ result in a lower $\mathcal{R}_{\rm{BBH}}(z)$, especially for large values of $\alpha_{\rm CE}$. In contrast, lower values of $f_{\rm MT}$ lead to a higher $\mathcal{R}_{\rm{BHNS}}(z)$. Finally, the impact on BNS merger rate density is very mild and depends on $\alpha_{\rm CE}$.

The physical reason is that highly non-conservative mass accretion significantly reduces the total mass of the binary star. In particular, the secondary star accretes just a small fraction of the mass lost by the primary star during Roche lobe overflow. This implies that non-conservative mass transfer enhances the formation of unequal mass binary compact objects, such as BHNSs. %, while conservative mass transfer tends to equalise the masses of the two compact objects.

%The effect of $f_{\rm MT}$ depends also on the value of $\alpha_{\rm CE}$ CE parameter.  We can say that generally the merger rate density tends to be higher when $f_{\rm MT} < 0.5$ for BBH and the opposite is given for BNS. {\filippo{This is expected since...}}

\subsection{Initial mass function}

%%%%%%%%%%%%%%%%%%FIGURE%%%%%%%%%%%%%%%%
\begin{figure}
\centering
\includegraphics[width=\columnwidth]{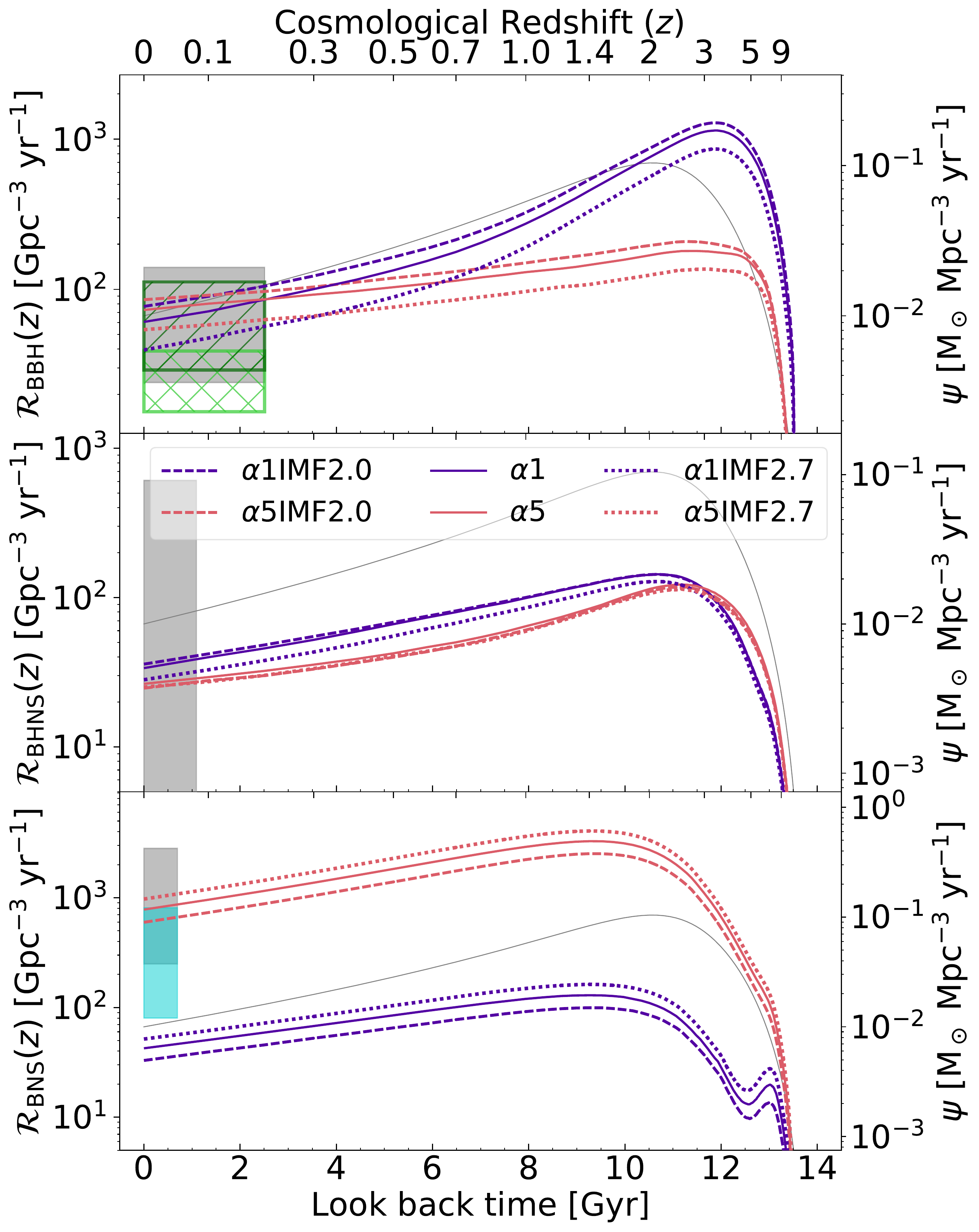}
\caption{Merger rate density of BBHs (top), BHNSs (centre) and BNSs (bottom).   Same as Figure~\ref{fig:mrd}, but we compare different values of the IMF slope $\alpha_{\rm IMF}$.}
\label{fig:imf}
\end{figure}
%%%%%%%%%%%%%%%%%%FIGURE%%%%%%%%%%%%%%%%

Figure~\ref{fig:imf} shows that  the impact of varying the IMF's slope on the cosmic merger rate is very mild, as already found by \cite{klencki2018}.
%Figure \ref{fig:imf} shows the impact of varying the IMF on the cosmic merger rate density evaluated with different slopes of the high-mass IMF. We notice that the IMF has a negligible impact on cosmic merger rate density, as it has been also found by \cite{klencki2018}. %{\filippo{check again their paper, maximum difference for BBH, check the similarties with that work, and also chruslinska has made something with IMF}}.
$\mathcal{R}_{\rm{BBH}}(z)$ and $\mathcal{R}_{\rm{BNS}}(z)$ show an opposite trend: the former is higher when a shallower IMF slope is considered. This result has a trivial explanation: if $\alpha_{\rm IMF} = 2.0$,  the fraction of massive stars that end up collapsing into BHs is higher with respect to $\alpha_{\rm IMF} = 2.7$.
%can be trivially explained by the trivial fact that $\alpha_{\rm IMF} = 2.0$ means that the fraction of massive stars that then can leave a merging BH is higher with respect to $\alpha_{\rm IMF} = 2.7$. As a result, the difference is maximum for the former compact binary class. The variation of the high-mass IMF does not affect the cosmic merger rate density by more than a factor of $\sim 2$. This can be explained by the fact that any change of the IMF slope demands a change on the SFR normalisation which at the end counterbalances the impact of high-mass IMF on the cosmic merger rate density \citep{chruslinska2020}. 

\subsection{Metallicity and SFR evolution}
\label{sec:metandsfr}

%%%%%%%%%%%%%%%%%%FIGURE%%%%%%%%%%%%%%%%
\begin{figure*}
\centering
\includegraphics[width=2\columnwidth]{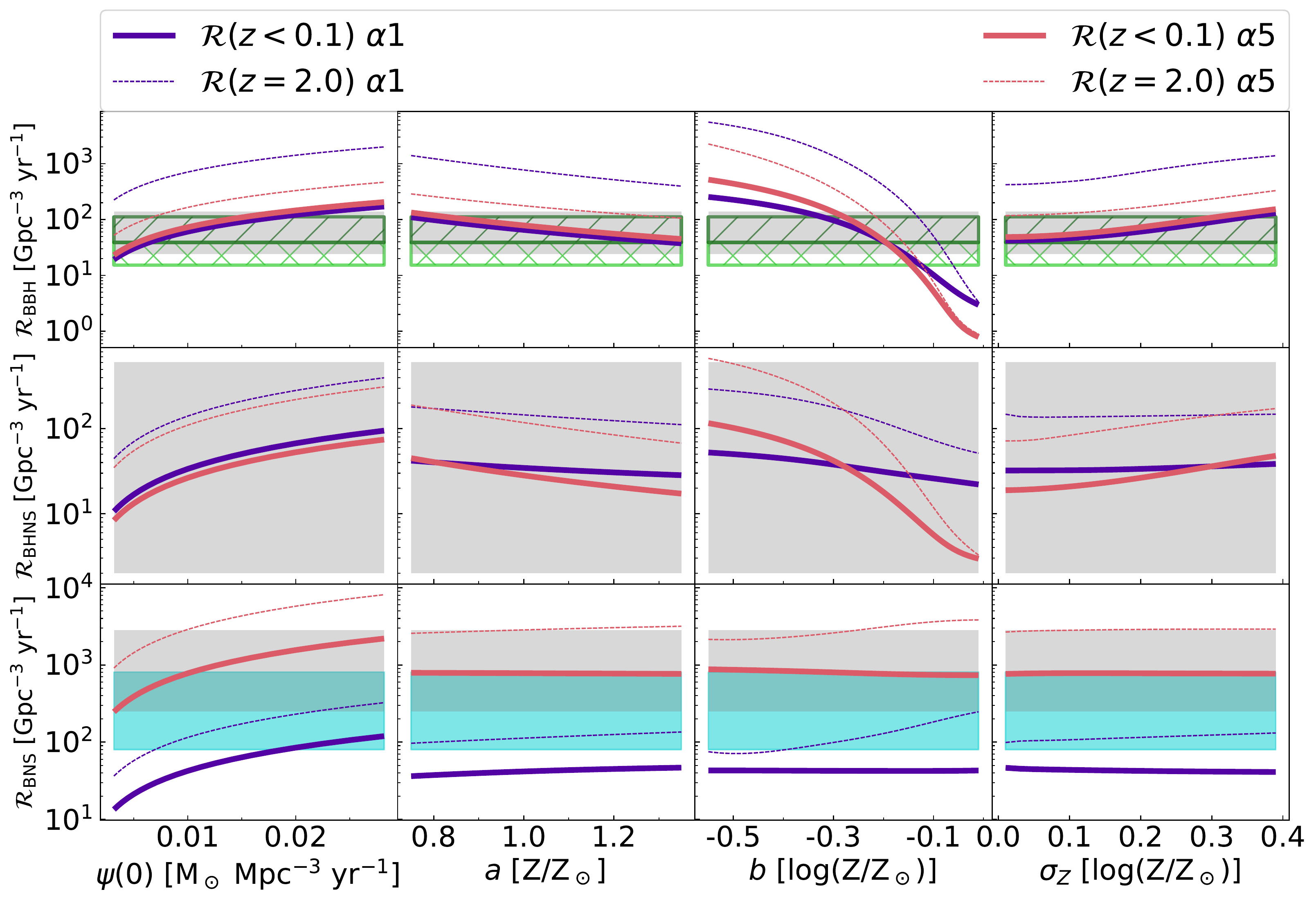}
\caption{Merger rate density in the local Universe $\mathcal{R}(z<0.1)$ (thick line) and at $z=2$ (thin dashed line) as a function of the SFR density normalisation $\psi{}(0)$ (equation~\ref{eq:sfrd}, leftmost column), the intercept $a$ and slope $b$ of the metallicity evolution model (equation~\ref{eq:metmod}, two central columns), and the metallicity spread $\sigma_{Z}$ (equation~\ref{eq:pdf}, rightmost column) for two different population-synthesis models: $\alpha1$ and $\alpha5$, as displayed in Table \ref{tab:models}. 
For BBHs and BHNSs, the grey shaded area shows the 90\% credible interval of the merger rate density in the local Universe, as inferred from the first two observing runs of the LVC. For BBHs, we consider the union of the rates obtained with model A, B and C in \protect\cite{abbottO2, abbottO2popandrate}. For BNSs, the grey shaded area shows the merger rate density estimated in \protect\cite{abbottGW190425}. The hatched green areas show the local BBH merger rate density inferred including O3a events \protect\citep{abbottpopO3a}. In particular, the dark-green and light-green hatched areas show the 90\% credible interval calculated including and excluding GW190814-like events, respectively \protect\citep{abbottpopO3a}. Finally, the cyan shaded area shows the 90\% credible interval of the local BNS merger rate density, as estimated by \protect\cite{abbottpopO3a}.}
\label{fig:everyunc}
\end{figure*}
%%%%%%%%%%%%%%%%%%FIGURE%%%%%%%%%%%%%%%%

%%%%%%%%%%%%%%%%%%FIGURE%%%%%%%%%%%%%%%%
\begin{figure}
\centering
\includegraphics[width= \columnwidth]{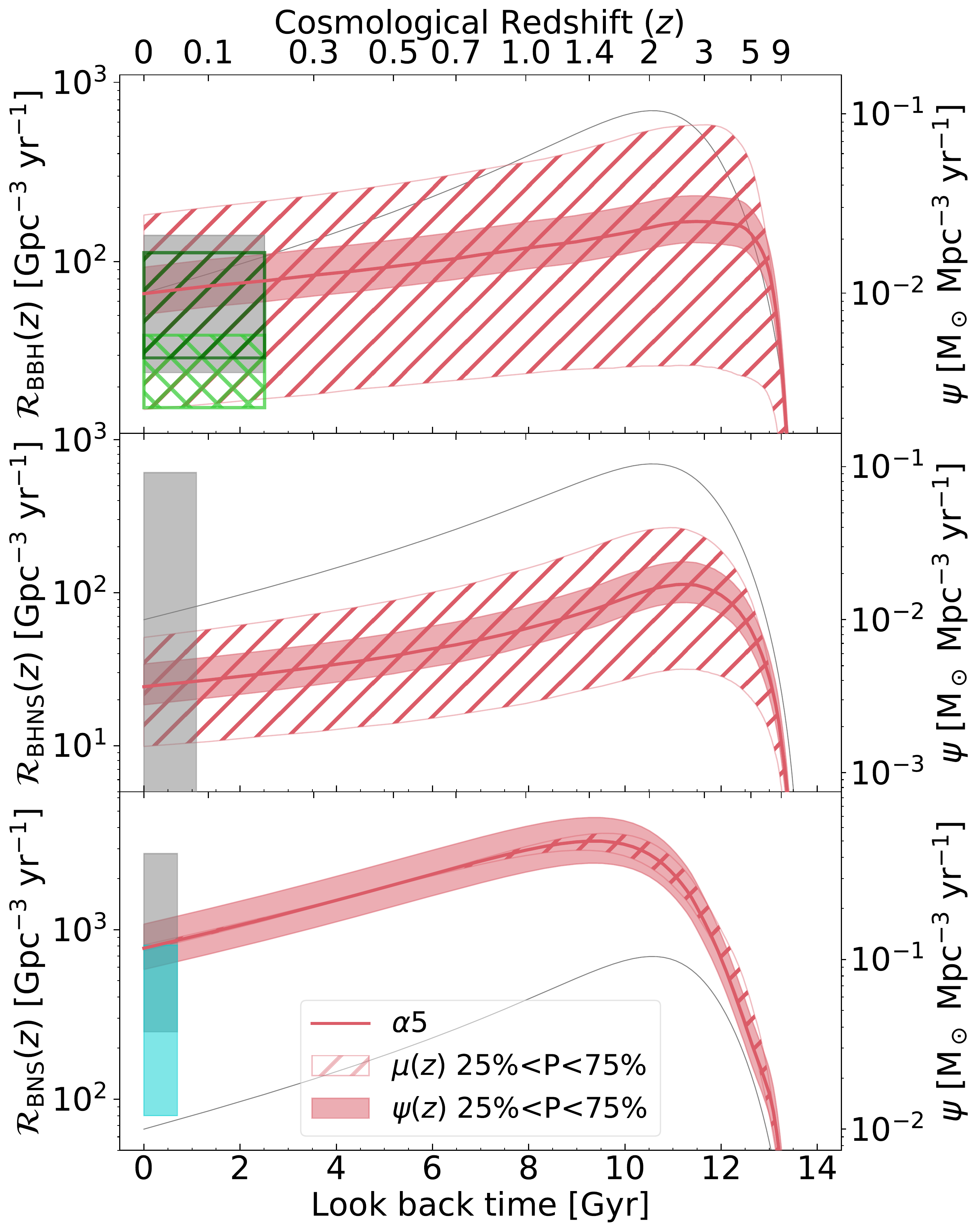}
\caption{
Merger rate density of BBHs (top), BHNSs (centre) and BNSs (bottom). Same as Figure~\ref{fig:mrd}, but we show the uncertainties on SFR and metallicity evolution. The contour areas represent 50\% of different realisations (between the 25\% and 75\% percentile), while the thick solid line is the median. See Section~\ref{sec:mrd} for details. To obtain the hatched area (with vertical lines), we varied only the slope and intercept of the metallicity fit (equation~\ref{eq:metmod}). To derive the shaded area we varied only the SFR density normalisation $\psi(0)$ (equation~\ref{eq:sfrd}). Hence, the hatched area and the shaded area quantify the uncertainty on metallicity and SFR, respectively.}
\label{fig:err}
\end{figure}
%%%%%%%%%%%%%%%%%%FIGURE%%%%%%%%%%%%%%%%

As we detailed in Section~\ref{sec:mrd}, the cosmic merger rate density is evaluated by assuming the fit from \cite{madau2017} for the SFR density (equation~\ref{eq:sfrd}) and a metallicity evolution model (equation~\ref{eq:metmod}). These two functions are affected by observational uncertainties; in this Section, we show their impact on the merger rate density. We take in account the uncertainty on four quantities, namely the normalisation factor of the SFR density $\psi(0)$ in equation~\ref{eq:sfrd}, the intercept $a$ and slope $b$ of equation~\ref{eq:metmod}, and the metallicity spread $\sigma_{Z}$ in equation~\ref{eq:pdf}. We assume the metallicity spread $\sigma_Z$ to follow a log-normal distribution with standard deviation $0.1$ dex.

We evaluate the cosmic merger rate density by varying the value of the aforementioned parameters in a [$-2\sigma,+2\sigma$] interval, where $\sigma$ is the standard deviation associated with each parameter. We assume here, for simplicity, that the considered quantities follow a Gaussian distribution and that they are not correlated with each other. %{\micmap{%The only parameter for which we did not assume an uncertainty distribution is the dispersion in the metallicity-redshift relation  $\sigma_{Z}$. In order to take into account of this uncertainty source, we vary the value of $\sigma_{Z}$ from 0.05 to 1 dex. 
%MM: cio\`e lo fai variare uniformemente? hmmm...per consistenza io lo farei variare anche lui su una gaussiana. Gaussiana con 0.05 dex?}}

Figure \ref{fig:everyunc} shows the dependence of the merger rate density on these observational parameters. For sake of clarity, we just plotted the merger rate density in the local Universe ($z_{\rm loc}<0.1$) and at $z=2$ for two different values of $\alpha_{\rm CE}$. 
$\mathcal{R}_{\rm{BNS}}(z)$ is only mildly affected by the parameters that concern metallicity ($a$, $b$ and, $\sigma_{Z}$), especially at low redshift. The most important parameter for BNSs is the normalisation of the SFR $\psi{}(0)$.  
In order for the local merger rate density to be within the 90\% credible interval inferred from the  O1, O2 and O3a GW data collection, we have to assume a value of $ \psi{}(0) \leq{} 0.01~\rm{M}_\odot$ Mpc$^{-3}$ yr$^{-1}$  ($\psi{}(0) \geq{} 0.02~\rm{M}_\odot$ Mpc$^{-3}$ yr$^{-1}$) for the model $\alpha{}5$ ($\alpha{}1$).
Thus, the cosmic merger rate density of BNSs is mainly affected by population-synthesis uncertainties and by the uncertainty on the SFR. %This result will be stressed again in Figure \ref{fig:err}.

In contrast, $\mathcal{R}_{\rm{BBH}}(z)$ changes by orders of magnitude when varying the parameters that describe metallicity evolution. 
For instance, if we assume $\sigma_{Z} > 0.35$ ($0.29$) dex for model $\alpha1$ ($\alpha5$), while keeping the other parameters at their fiducial values, the local merger rate density of BBHs is outside the 90\% credible interval inferred by the LVC from the GWTC-2, evaluated %onsidering in the analysis the secondary mass to be as low as $m_{2}>2$M$_\odot$ i.e. 
including GW190814-like events. 
We expect $\mathcal{R}_{\rm BBH}$ to grow with $\sigma_{Z}$ 
because a larger value of $\sigma_{Z}$ means that  the percentage of metal-poor stars at low redshift is higher. As we have seen from Figure \ref{fig:eta}, the BBH merger efficiency is orders of magnitude higher for metal-poor stars. For the same reason, the cosmic merger rate density of BBHs decreases for increasing values of the intercept in equation~\ref{eq:metmod}.

The value of the slope $b$ in equation~\ref{eq:metmod} represents the largest source of uncertainty for $\mathcal{R}_{\rm BBH}$, compared to the other observational parameters. The local BBH merger rate density changes by two to four orders of magnitude by varying $b$ within 2 $\sigma{}$. 
The local merger rate density is inside the 90\% credible interval inferred from GWTC-2 only for $b\in{}[ -0.19, -0.12]$ ($[ -0.19, -0.15]$) for the model $\alpha1$ ($\alpha5$).

BHNSs behave in a similar way to BBHs, but all the considered realisations are still within the upper limit from the LVC.

Figure \ref{fig:err} shows the overall uncertainty affecting the cosmic merger rate density due to SFR and metallicity. %affects the metallicity evolution model (Equation~\ref{eq:metmod}) and the cosmic SFR  density (Equation~\ref{eq:sfrd}). 
We evaluate this uncertainty through the Monte Carlo method presented in  Section~\ref{sec:mrd}. 
$\mathcal{R}_{\rm{BBH}}(z)$ and $\mathcal{R}_{\rm{BHNS}}(z)$  are heavily affected by uncertainties on metallicity evolution.  In contrast, the uncertainty on $\mathcal{R}_{\rm{BNS}}(z)$ is much smaller and is dominated by the SFR. 

\subsection{Merger rate density as a function of metallicity}
%%%%%%%%%%%%%%%%%%%%%%%%%%FIGURE%%%%%%%%%%%%%%%%%%%%%%%%%%%%%%%%%%%%%%
\begin{figure*}
\centering
\includegraphics[width=2\columnwidth]{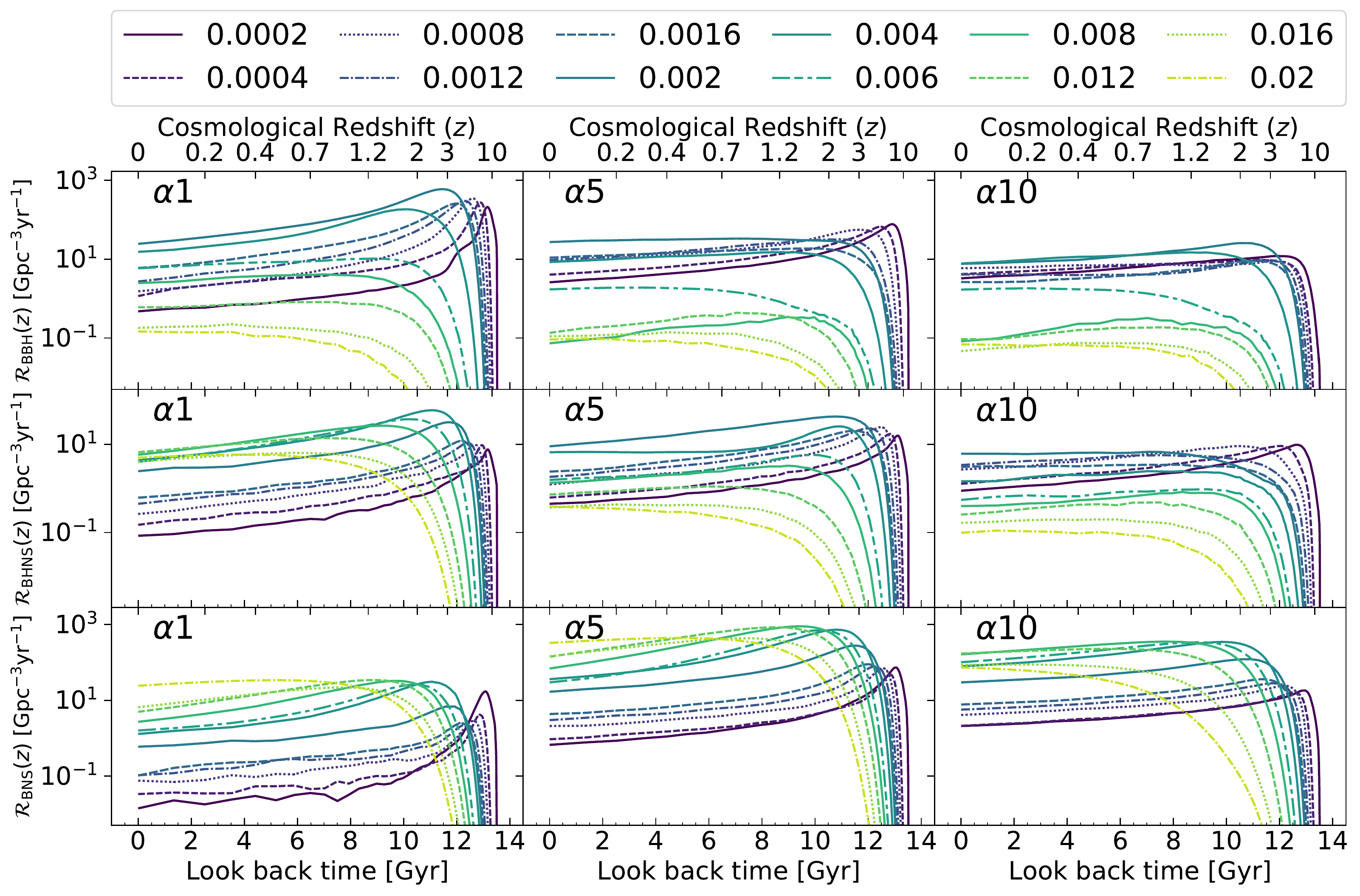}
\caption{Contribution of progenitor's metallicity  to the cosmic merger rate density for three different population-synthesis model: $\alpha1$, $\alpha5$, and $\alpha10$, as reported in Table \ref{tab:models}. The considered metallicities are shown in the legend in the upper panel ($Z=0.0002-0.02$).}
\label{fig:mrdZ}
\end{figure*}
%%%%%%%%%%%%%%%%%%%%%%%%%%%%%%%%%%%%%%%%%%%%%%%%%%%%%%%%%%%%%%%%%%%%%%

Figure \ref{fig:mrdZ} shows the contribution of different progenitor's metallicities to the cosmic merger rate density, for three different values of $\alpha_{\rm CE}=1,$ 5 and 10. For $\alpha_{\rm CE}=1$, progenitor stars with $Z\sim{}0.004$ 
produce most of the BBHs merging at $z\lesssim{}4$.

In contrast,  $\mathcal{R}_{\rm{BNS}}(z)$ is dominated by solar metallicity progenitors for $z\lesssim{}1$. Again, this  springs from the different dependence of BBH and BNS merger efficiency on metallicity.

Different values of $\alpha_{\rm CE}$ change the relative contribution of different metallicities to the merger rate. For all kind of compact object binaries considered here (BBHs, BHNSs and BNSs), larger values of $\alpha_{\rm CE}$ correspond to a larger contribution of metal-poor stellar populations to the local merger rate with respect to metal-rich stellar populations.  
This happens because the delay times are generally longer for large values of $\alpha_{\rm CE}$ than for small values of $\alpha_{\rm CE}$. In fact, larger values of $\alpha_{\rm CE}$ imply that the CE is ejected without much shrinking of the binary system. Hence, the final binary that emerges from CE has a larger orbital separation, and needs more time to merge by GW emission.

\section{Discussion}

\subsection{Fitting the merger rate density at \textit{z}<1}
\label{sec:fit}

Our models show that the merger rate density of binary compact objects is broadly reminiscent of the cosmic SFR density. Here, we want to quantify how close is the slope of the merger rate density to that of the cosmic SFR in our different models. Since LIGO and Virgo at design sensitivity will observe BBH mergers up to $z\sim{}1$, we restrict our attention to the slope of the merger rate density up to such redshift \citep{Fishbach2018}. We assume that $\mathcal{R}_{\rm BBH}(z)\propto{}(1+z)^\lambda{}$  if $z<1$. Under such assumption, we can fit the following quantities
\begin{equation}\label{eq:fit}
\log{ \left[\mathcal{R}(1+z)\right]} = \log{ \mathcal{R}_0 } + \lambda \,{}\log{(1+z)}.
\end{equation}
We expect to find $\lambda{}\approx{}2.6$ if the merger rate density scales approximately with the cosmic SFR density, given equation~\ref{eq:sfrd}.

%%%%%%%%%%%%%%%%%%TABLE%%%%%%%%%%%%%%%%
\begin{table}
    \caption{Coefficients of the  fit in equation~\ref{eq:fit}  with $0\leq{}z<1$ for each considered model. %{\micmap{MM: I am wondering whether it is possible to complement this table with a figure, e.g. y axis $\lambda$, x-axis model name.}}{\yann{I totally agree}}
    } 
    \begin{center}
    \begin{tabular}{l c c c c c c}
    \toprule
Model name & \multicolumn{2}{c}{BBH} & \multicolumn{2}{c}{BHNS} & \multicolumn{2}{c}{BNS} \\
\midrule
& $\mathcal{R}_0$ & $\lambda$ & $\mathcal{R}_0$ & $\lambda$ & $\mathcal{R}_0$ & $\lambda$ \\
\midrule
$\alpha$0.5	& 40.39	& 1.56	& 14.74	& 2.12	& 21.75	& 1.04\\ 
$\alpha$1	& 57.74	& 2.16	& 34.45	& 1.57	& 44.59	& 1.50\\ 
$\alpha$2	& 105.42	& 1.33	& 29.58	& 1.60	& 76.54	& 1.95\\ 
$\alpha$3	& 94.08	& 1.19	& 30.10	& 1.47	& 358.44	& 1.97\\ 
$\alpha$5	& 73.76	& 0.83	& 25.74	& 1.28	& 812.20	& 1.92\\ 
$\alpha$7	& 52.08	& 0.74	& 26.88	& 1.00	& 1036.82	& 1.29\\ 
$\alpha$10	& 37.09	& 0.77	& 20.23	& 0.72	& 746.12	& 0.84\\ \midrule 
$\alpha$1s265	& 12.36	& 2.13	& 2.07	& 2.26	& 4.26	& 2.01\\ 
$\alpha$5s265	& 10.00	& 1.97	& 1.84	& 2.02	& 39.17	& 2.61\\ 
$\alpha$1s150	& 36.16	& 1.86	& 8.10	& 2.07	& 12.83	& 1.72\\ 
$\alpha$5s150	& 31.69	& 1.71	& 8.32	& 1.73	& 124.97	& 2.38\\ 
$\alpha$1s50	& 57.78	& 1.82	& 41.62	& 1.70	& 59.85	& 1.53\\ 
$\alpha$5s50	& 74.93	& 1.03	& 36.20	& 1.16	& 544.78	& 2.05\\ 
$\alpha$1F12	& 53.82	& 1.96	& 9.73	& 1.89	& 7.67	& 1.87\\ 
$\alpha$5F12	& 49.85	& 0.94	& 6.22	& 1.76	& 67.94	& 2.50\\ 
$\alpha$1VG18	& 54.07	& 1.95	& 9.61	& 1.89	& 17.31	& 1.93\\ 
$\alpha$5VG18	& 49.69	& 0.94	& 6.57	& 1.72	& 147.00	& 2.46\\ \midrule 
$\alpha$1R	& 65.28	& 2.52	& 42.54	& 1.31	& 35.31	& 1.41\\ 
$\alpha$5R	& 95.73	& 0.35	& 12.34	& 2.01	& 669.05	& 1.75\\ \midrule 
$\alpha$1MT0.1	& 83.03	& 1.71	& 128.04	& 2.05	& 82.99	& 1.77\\ 
$\alpha$5MT0.1	& 35.95	& 0.78	& 110.34	& 1.58	& 384.90	& 2.31\\ 
$\alpha$10MT0.1	& 11.39	& 0.31	& 59.69	& 0.48	& 464.31	& 1.42\\ 
$\alpha$1MT0.5	& 71.20	& 2.10	& 59.29	& 2.05	& 60.13	& 1.71\\ 
$\alpha$5MT0.5	& 77.70	& 0.43	& 37.36	& 0.96	& 535.59	& 2.36\\ 
$\alpha$10MT0.5	& 37.50	& 0.25	& 15.35	& 0.49	& 836.07	& 1.36\\ 
$\alpha$1MT1.0	& 58.47	& 2.13	& 34.06	& 1.58	& 43.88	& 1.53\\ 
$\alpha$5MT1.0	& 73.18	& 0.84	& 25.67	& 1.27	& 813.39	& 1.93\\ 
$\alpha$10MT1.0	& 36.99	& 0.78	& 20.55	& 0.69	& 761.75	& 0.90\\ \midrule 
$\alpha$1IMF2.0	& 72.98	& 2.07	& 36.70	& 1.49	& 34.60	& 1.48\\ 
$\alpha$5IMF2.0	& 84.34	& 0.83	& 24.07	& 1.29	& 620.73	& 1.92\\ 
$\alpha$1IMF2.7	& 37.45	& 2.25	& 28.28	& 1.62	& 53.97	& 1.55\\ 
$\alpha$5IMF2.7	& 53.71	& 0.86	& 24.36	& 1.26	& 1013.70	& 1.92\\ 
 \bottomrule
    \end{tabular}
    \end{center}
    	\footnotesize{$\mathcal{R}_0$ is given in [Gpc$^{-3}$ yr$^{-1}$]. In order to check the goodness of the fits, we calculated the \textit{coefficient of determination} $\rm R^2$ which is $>0.95$ for all the linear fits, except for the model $\alpha$5MT0.1, which yields $\rm R^2 = 0.84$. %Thus, we consider all these fits robust.
    	}
    \label{tab:fit}
\end{table}

%%%%%%%%%%%%%%%%%FIGURE%%%%%%%%%%%%%%%%%%%%%%%%
\begin{figure*}
\centering
\includegraphics[width=2\columnwidth]{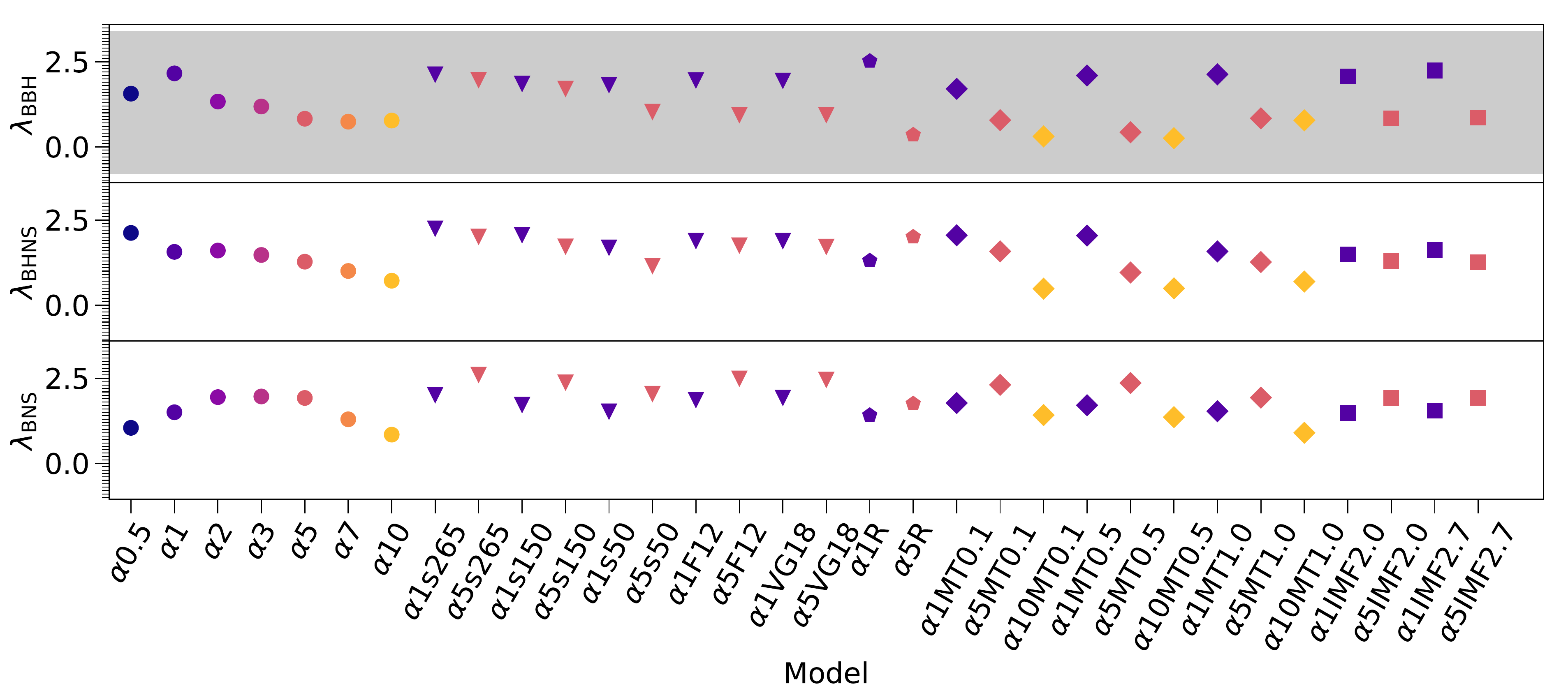}
\caption{ Values of $\lambda$ (defined in equation~\ref{eq:fit}) for each considered model. From top to bottom: BBHs, BHNSs and BNSs. The grey shaded area represents the 90\% credible interval of $\lambda{}_{\rm BBH} = 1.3^{+2.1}_{-2.1}$, inferred by the LVC from GWTC-2, adopting the {\sc Power
Law + Peak model} \citep{abbottpopO3a}. 
}
\label{fig:fit}
\end{figure*}
%%%%%%%%%%%%%%%%%%%%%%%%%%%%%%%%%%%%%%%%%%%

We show the results of the fit  for $z\in[0,1)$ in Table~\ref{tab:fit} and Figure \ref{fig:fit}. Most of our models have $\lambda{}<2.6$ for BBHs, BHNSs and BNSs. This suggests that the actual slope of the merger rate density is shallower than the one of the cosmic SFR, because of the delay time distribution, which encodes information on binary evolution processes, and because of the impact of metallicity on the merger efficiency.

The model closest to $\lambda =2.6$ is $\alpha5$s265 for BNSs, i.e. the model with large natal kicks. With this kick choice, only the tightest and most massive systems can survive the SN explosion, and these systems are also those that merge with the shortest delay times by GW radiation. In contrast, the model with the shallowest slope is $\alpha10$MT0.5 for BBHs, which yields $\lambda{}=0.25$. As we have already seen in Figure~\ref{fig:mrdZ}, models with $\alpha_{\rm CE}=10$ have longer delay times than the other models.
%From Section \ref{sec:mock}, we have seen that the cosmic merger rate is closer to the SFR when the merger efficiency is constant and the delay time distribution is peaked in the Myr range values. In fact, $\alpha5$s265 is a particular case for BNSs. With this parameter choice, only the closest and most massive systems can survive the SN explosion, and this type of systems are also those that merge earlier by GW radiation. \\

%A similar behaviour is shown by BHNS with $\alpha0.5$. From Figure \ref{fig:eta}, we notice that the merger efficiency of this case is almost independent with  $Z$. We can compare this result with Figure \ref{fig:mock}, where a constant $\eta$ has been also taken into account. Since the leading slope is $\lambda = 2.12$, we are allowed to assume that the delay time distribution is also peaked in the Myr range.

%On the other hand, 

%In this case, $\lambda = 0.25$ might be attributed to an almost flat distribution of the delay times and a BBH-like merger efficiency, as we have seen from Figure \ref{fig:mock}, bottom panel. \\

%Therefore,
%This analysis wants to highlight the feasibility to connect the  properties of the delay time distribution with the leading slope value $\lambda$ in equation~\ref{eq:fit}, once we set the typical merger efficiency evolution with metallicity of that compact binary type. We stressed the importance on focusing on the delay time distribution of a population of merging compact binaries, since it contains all the information coming from binary astrophysics.  

The slope $\lambda$ of each of considered model in this work is within the 90\% credible interval inferred by the LVC \citep{abbottpopO3a} for BBHs, as also shown in Figure \ref{fig:fit}.

\subsection{Merger efficiency and delay time impact on merger rate density}
\label{sec:mock}

%%%%%%%%%%%%%%%%%FIGURE%%%%%%%%%%%%%%%%%%%%%%%%
\begin{figure}
\centering
\includegraphics[width= \columnwidth]{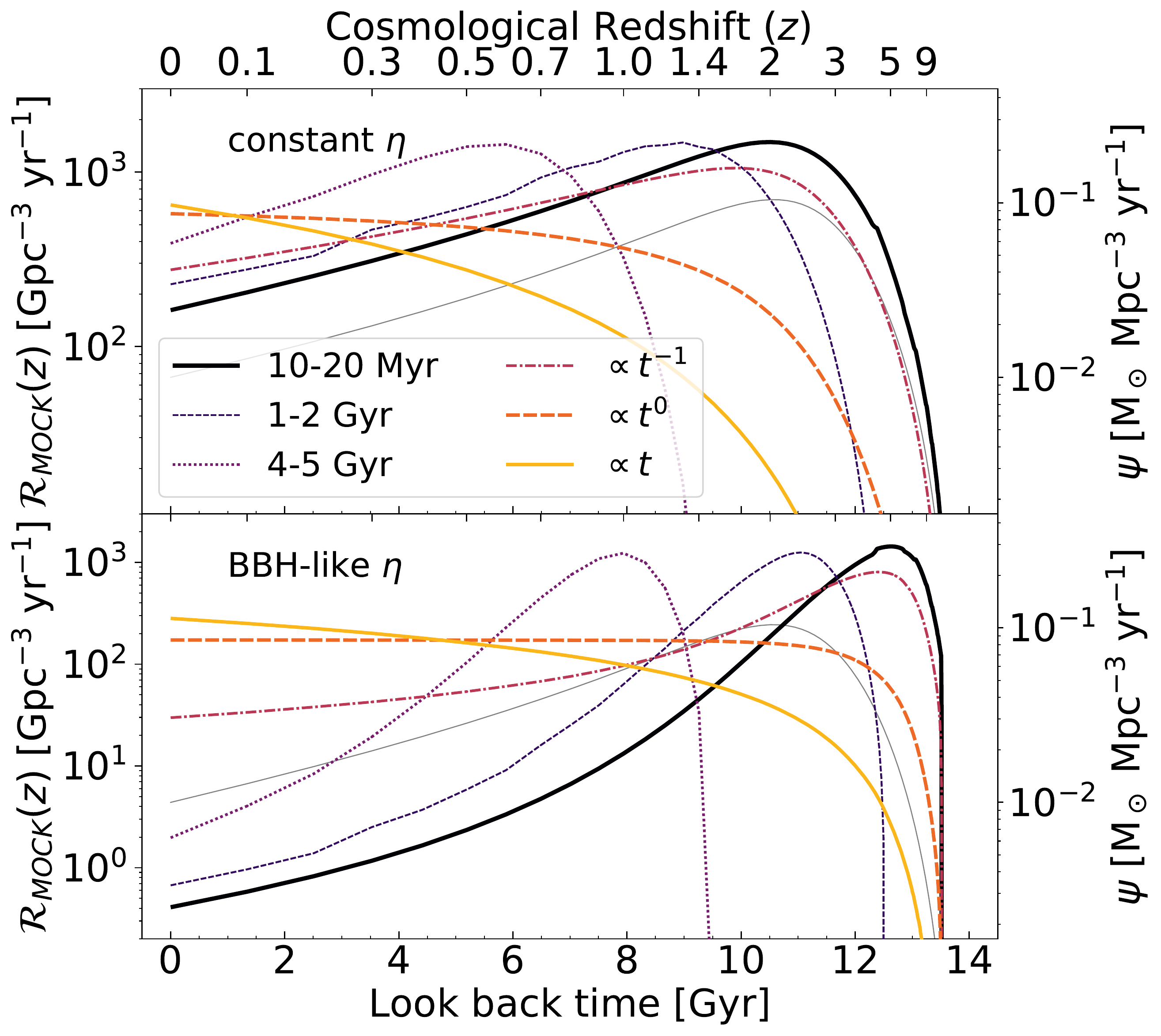}
\caption{Evolution of the cosmic merger rate density evaluated with mock catalogues of merging compact binaries with constant (top) and BBH-like (bottom) merger efficiency $\eta$; and with six different delay time distributions. The thin grey line is the SFR from \protect\cite{madau2017}.}
\label{fig:mock}
\end{figure}
%%%%%%%%%%%%%%%%%%%%%%%%%%%%%%%%%%%%%%%%%%%

In this Section, we want to use a simple toy model to interpret the results we found in the previous Section. In order to understand what are the effects on the cosmic merger rate density of the convolution of the SFR density with different delay time distributions and with metallicity, we performed some mock simulations\footnote{Mock models have been extensively applied in the early years of binary evolution studies, where they were adopted to interpret short gamma-ray burst redshift distributions (see for instance \citealt{nakar2006, zheng2007, berger2007}).}.

The first ingredient of our mock simulations is the merger efficiency, which encodes a possible dependence on metallicity. We consider two different cases. In the first case, we assume a constant merger efficiency $\eta$, independent of metallicity; in the second case, we adopt a BBH-like $\eta$, higher at low metallicity. Specifically, for the latter case we use the merger efficiency of the $\alpha5$ case, as displayed in Figure \ref{fig:eta}. 

The second ingredient is the %At this point, in order to evaluate the merger rate density, we had to assume a 
delay time $t_{\rm del}$ distribution.  For simplicity, we assume that the delay time distribution does not depend on  metallicity. We consider four cases in which we assume a uniform delay time distribution: three narrow distributions with $t_{\rm del}$ uniform from  10 to 20 Myr, from 1 to 2 Gyr and from 4 to 5 Gyr; and a broader distribution with $t_{\rm del}$ uniform from 10 Myr to 14 Gyr. Then, we consider two power law distributions: $\propto{}t^{-1}$ and $\propto{}t$, defined from 10 Myr to 14 Gyr. 

Figure \ref{fig:mock} shows the merger rate density evaluated with the aforementioned mock simulations. %If we look at the models with constant merger efficiency, we see how the case where we made use of the narrowest delay time distributions better reproduces the SFR density. Basically 
Let us start considering the cases with constant $\eta{}$. If the delay time is uniformly distributed between 10 and 20 Myr, the merger rate density has exactly the same slope and peak redshift as the cosmic SFR. The other two narrow delay time distributions have the effect to shift the merger rate density peak towards lower redshifts than the peak of the cosmic SFR. The case with $dN/dt\propto{}t^{-1}$ has a very similar slope to the cosmic SFR density ($\lambda\sim{}2.6$), while the cases with $dN/dt\propto{}t^0$ and $t$ have significantly flatter and even upturning slopes ($\lambda<0$).
The case with constant $\eta$ and $\propto{}t^{-1}$ delay time distribution is reminiscent of our BNS simulations. However, the fact that our BNS models generally have a slope flatter than $\lambda{}=2.6$ (Table~\ref{tab:fit}) tells us that, for a constant $\eta{}$, the  delay time distribution in our models is flatter than $t^{-1}$.

%It is worth noting that the uniform and $t^{+1}$ distributions yield a merger rate density that decreases with redshift, while $t^{-1}$ provide a merger rate density which has an evolution of the same SFR-density shape, i.e. it shows similar but a lower leading slope. This latter case, constant $\eta$ and $t^{-1}$ delay time distribution, is the one which is more representative of the BNS case.

Let us now look at the cases with a BBH-like $\eta{}$. %case also presents interesting features. We notice that 
The delay time distribution uniform between 10 and 20 Myr peaks at a higher redshift ($z_{\rm{peak}}\gtrsim{}5$) with respect to the cosmic SFR density. This happens because the BBH-like merger efficiency is maximum for metallicity $Z\sim{}0.0002$, which is common in the early Universe. This result is similar to our BBH models with $\alpha_{\rm CE}\leq{}1$ and is indicative of a strong dependence on metallicity combined with short delay times. %the SFR evolution is weighted and quenched by the merger efficiency which is at least two orders of magnitude lower at solar metallicity and so at lower redshift, with respect to higher metallicities. 
%The other two narrow delay time distributions produce the same effect as in the case with constant $\eta{}$: they shift the merger rate density peak towards lower redshifts. The case with $t^{+1}$ decreases with redshift, while 
For a uniform delay time distribution between 10 Myr and 14 Gyr, the merger rate density is almost constant with time, similar to the trend of $\mathcal{R}_{\rm BBH}(z)$ in the $\alpha10$ model. %This latter case is very interesting especially if compared with Figure \ref{fig:mrd}, top panel. We see that for instance $\mathcal{R}_{\rm{BBH}}(z)$ evaluated with $\alpha10$ mildly increases with increasing redshift; 
%This is a clear clue that the delay time distribution is almost flat (Giacobbo et al. 2020, in preparation). 
Indeed, the delay time distribution of $\alpha10$  is nearly flat, because $\alpha_{\rm CE} > 5$ implies less effective shrinking of the binaries during CE, hence longer delay times.

\subsection{Comparison with previous work}
%{\micmap{ BISOGNA ELABORARE O TOGLIERE We present a comparison between the result obtained with our semi-analytic model and with the Illustris and the EAGLE cosmological simulations in Appendix \ref{appendix}. }}

One of the main results of our analysis is that the BBH merger rate varies by more than one order of magnitude because of uncertainties on metallicity evolution, while the merger rate of BNSs is substantially unaffected by metallicity. This result is in agreement with previous studies (e.g. \citealt{giacobbo2018b,chruslinska2019,neijssel2019,tang2020,belczynski2020}).

On top of that, the merger rates of BBHs, BHNSs and BNSs strongly depend on CE efficiency ($\alpha_{\rm CE}$), mass transfer efficiency ($f_{\rm MT}$) and  natal kicks. For the merger rate of BBHs, the uncertainty connected with such binary evolution parameters is of the same order of magnitude as the uncertainty on metallicity evolution, consistent with \cite{neijssel2019} and \cite{belczynski2020}. In these previous studies, the authors pointed out the importance of inter-parameter degeneracy while deriving astrophysical conclusions from GW  observations.

For a suitable choice of these binary evolution parameters (namely $\alpha_{\rm CE}\ge{}2$ and  moderately low natal kicks), 
we find reasonable agreement between our models and the LVC rates  after O1, O2 and O3a \citep{abbottO2,abbottO2popandrate,abbottGW190425,abbottpopO3a}. In particular, only models with moderately low kicks (depending on the ejected mass and the SN model), such as these described by  \cite{giacobbo2020}, \cite{tang2020}, \cite{zevin2020} and \cite{vignagomez2018}, can match the BNS merger rate in the local Universe.

\begin{figure}
\centering
\includegraphics[width=\columnwidth]{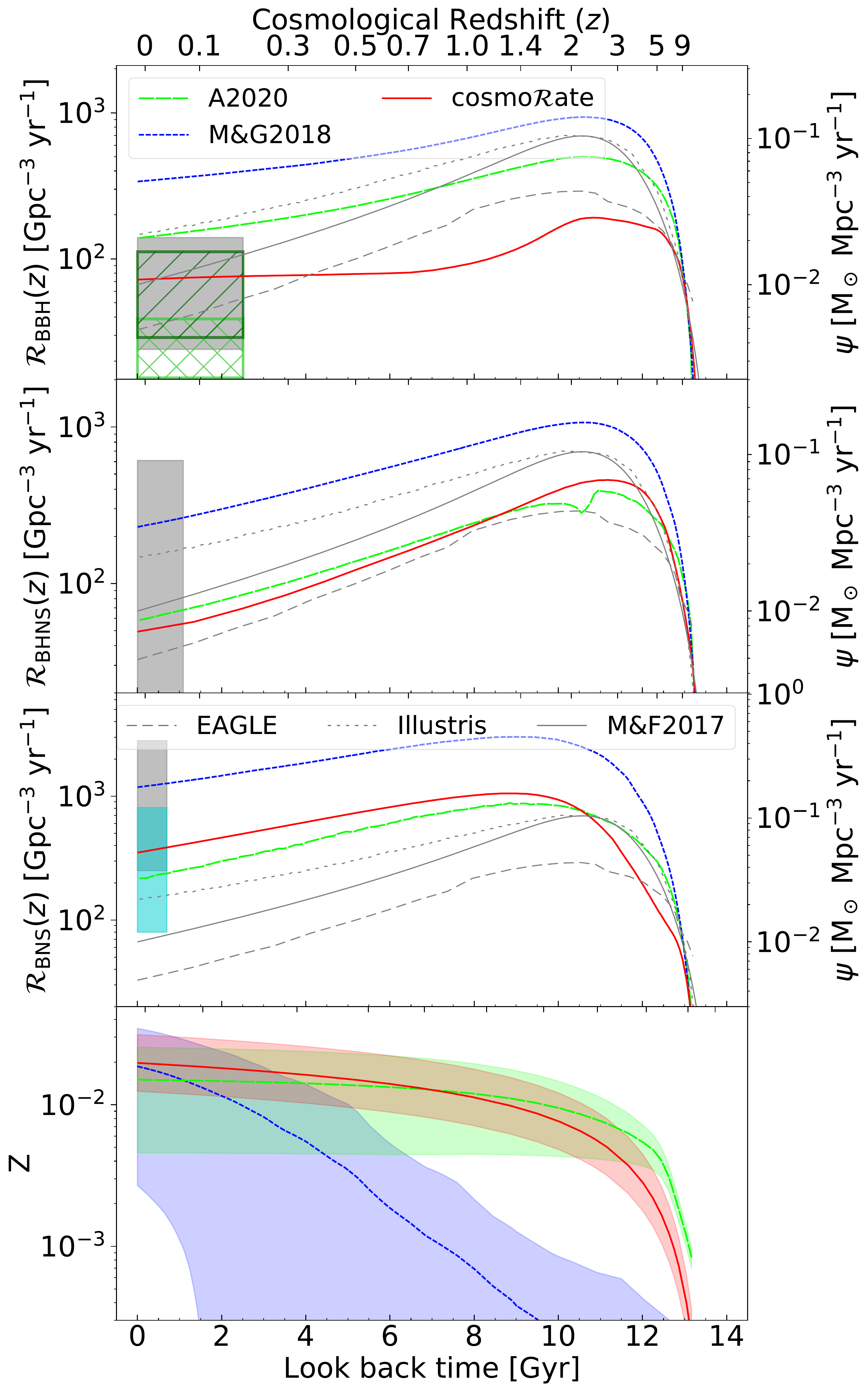} 
\caption{Comparison between the merger rate density of BBHs, BHNSs and BNSs obtained with {\sc cosmo$\mathcal{R}$ate}  (red solid lines) and the ones derived by \protect \citet[] [green long-dashed lines with label A2020]{artale2020}, based on the {\sc eagle} cosmological simulation \protect \citep{eagle2015}, and by \protect\citet[][blue short dashed lines with label M\&G2018]{mapelli2018}, based on the {\sc illustris} cosmological simulation \protect\citep{vogelsberger2014a}. Solid grey line: SFR density from \protect\citet[][with label M\&F2017]{madau2017}; long-dashed grey line: SFR density from the {\sc eagle} simulation; short-dashed grey line: SFR density from the {\sc illustris} simulation. The lower panel shows the average metallicity evolution in each model. Short-dashed blue line: {\sc illustris}; long-dashed green line: {\sc eagle}; solid red line: this work. The shaded area shows one standard deviation from the average metallicity. 
%encloses the region with metallicity within $\pm 1\sigma$, where $\sigma$ is the standard deviation of the metallicity distribution model at each given redshift.}
}
\label{fig:cosmosim}
\end{figure}

When we compare our results with models adopting the cosmic SFR and metallicity evolution from cosmological simulations \citep[e.g.,][]{lamberts2016,mapelli2017,mapelli2018,artale2020}, we find more conspicuous differences. For example, Figure~\ref{fig:cosmosim} shows the comparison between the merger rates estimated with {\sc cosmo$\mathcal{R}$ate} and those estimated by \cite{mapelli2018} and \cite{artale2020}, using the {\sc illustris} \citep{vogelsberger2014a,vogelsberger2014b,nelson2015} and the {\sc eagle} cosmological simulation \citep{eagle2015}, respectively.
%%%%%%%%%%%%%%%%%%%%%FIGURE%%%%%%%%%%%%%%%%%%%%%%%%%%%%%%
%%%%%%%%%%%%%%
To make a one-to-one comparison, we have re-run {\sc cosmo$\mathcal{R}$ate} with the binary compact object catalogues from model  CC15$\alpha5$, obtained with an old version of {\sc mobse} \citep[see][]{giacobbo2018b} and adopted in both \cite{mapelli2018} and \cite{artale2020}. The merger rate density of BBHs, BHNSs and BNSs in the local Universe is a factor of $\sim{}3-5$ higher in \cite{mapelli2018} than in this work. This difference is due to the cosmic SFR of the {\sc illustris} cosmological simulation, which is a factor of $\sim{}2-2.5$ higher in the local Universe than the one described by \cite{madau2017}, and to the metallicity evolution of the {\sc illustris}, which has a larger contribution from metal-poor stars (see the bottom panel of Figure \ref{fig:cosmosim}). The results of {\sc{cosmo$\mathcal{R}$ate}} are more similar to those reported in \cite{artale2020}. However, the cosmic SFR of the {\sc eagle} is significantly lower than the one measured by \cite{madau2017}, as reported previously by \citet{Katsianis2017}. This is compensated by the fact that the {\sc eagle} average metallicity in the local Universe is lower with respect to equation~\ref{eq:metmod}.

\section{Summary}

We investigated the cosmic merger rate density evolution of compact binaries, by exploring the main sources of uncertainty. %the parameter space that affects the cosmic merger rate density of compact binaries. In order to achieve this, 
We have made use of the  {\sc{cosmo$\mathcal{R}$ate}} code \citep{santoliquido2020}, which evaluates the cosmic merger rate density by combining catalogues of merging compact binaries, obtained from population-synthesis simulations, with the \cite{madau2017} fit to the SFR density and with a metallicity evolution model based on \cite{decia2018} and \cite{gallazzi2008}. 

We took into account uncertainties on the most relevant binary evolution processes: CE, SN kicks, core-collapse SN models and mass accretion by Roche lobe overflow. These represent the main bulk of uncertainty on the merger rate density due to binary evolution prescriptions. In addition, we varied the slope of the IMF. Our results confirm that the core-collapse SN model and the IMF produce negligible variations of the merger rate density.

The parameter $\alpha_{\rm CE}$, quantifying the efficiency of CE ejection, is one of the main sources of uncertainty. The merger rate density of BNSs spans up to 2 orders of magnitude if $\alpha_{\rm CE}$ varies from 0.5 to 10. For the same range of $\alpha_{\rm CE}$, $\mathcal{R}_{\rm{BHNS}}(z)$ and $\mathcal{R}_{\rm{BBH}}(z)$  vary up to a factor of 2 and 3, respectively. Only values of $\alpha_{\rm CE}\ge{}2$  give local BNS merger rate densities within the 90\% credible interval inferred from GWTC-2 \citep{abbottO2,abbottO2popandrate,abbottGW190425} when we adopt our fiducial kick model.

 Large natal kicks ($\sigma{}=265$ km s$^{-1}$) yield BNSs merger rate densities below the 90\% credible interval from GWTC-2. Only models with moderately low kicks and large values of $\alpha_{\rm CE}$ \citep{bray2016,bray2018,vignagomez2018,giacobbo2020,tang2020} predict values of the BNS merger rate within the 90\% LVC credible interval inferred from GWTC-2.

%We highlighted that the lower is the SN kick mean velocity, the lower is the cosmic merger rate density. This is true for all compact binary classes, especially for BNSs.
%Whilst the difference on the cosmic merger rate density due to different SN kick velocity distributions can be considered negligible for BBHs and BHNSs, this is not the case for BNSs where the merger rate density can vary up to an order of magnitude.  

Different values of the mass transfer efficiency parameter do not result in appreciable differences in the BNS merger rate density. The difference between $f_{\rm MT}=0.1$ and $f_{\rm MT}=1$ (conservative mass transfer) is up to a factor of $5-10$ for BHNSs and BBHs. %However, it remains always under one order of magnitude. It is worth noting that
The BBH local merger rate density with $f_{MT} = 0.1$ can be as low as $\mathcal{R}_{\rm {BBH}}(z_{\rm{loc}}<0.1) \sim 11$ Gpc$^{-3}$ yr$^{-1}$ with $\alpha_{\rm CE} = 10$,  within the 90\% credible interval inferred from GWTC-2 \citep{abbottO3a,abbottpopO3a}.

%It is worth noting that 
\cite{callister2020} show that models with local merger rates  $\mathcal{R}_{\rm BBH}(z)\propto{}(1+z)^\lambda{}$ with $\lambda{}\ge{}7$ are rejected, based on the O1 and O2 LVC data and on the analysis of the stochastic background. Recently, the LVC has reported $\lambda=1.3^{+2.1}_{-2.1}$  within the 90\% credible interval, based on GWTC-2 \citep{abbottpopO3a}. All of our models yield a slope $\lambda \leq{} 2.6$ for $z<1$; hence, none of them can be rejected by current data. %Furthermore, Figure \ref{fig:fit} shows that our model's $\lambda$ is always within the 90\% credible interval reported in GWTC-2 \citep{abbottpopO3a}. 
Most of our models are fitted by $\lambda{}\leq{}2$, a shallower slope with respect to the cosmic SFR. We show that this is indicative of a delay time distribution flatter than $t^{-1}$. 

We have also investigated the effect of observational uncertainties on the cosmic SFR and on metallicity evolution. $\mathcal{R}_{\rm{BNS}}(z)$ is not significantly affected by metallicity evolution (Figure ~\ref{fig:err}).  %We assumed as main source of uncertainty from the SFR density the normalisation constant $\psi(0)$ (equation~\ref{eq:sfrd}). We have shown that in order for $\mathcal{R}_{\rm{BNS}}(z)$ to be consistent 
%The BNS local merger rate density is within the 90\% credible interval inferred from the  LVC, only if the current SFR density $\psi(0)$ is very high (beyond $+2\,{}\sigma$) or the CE ejection efficiency parameter is $\alpha_{\rm CE} \ge{} 3$.
In contrast, the metallicity evolution  has a tremendous impact on the merger rate
density of BBHs (Figure \ref{fig:err}). %It can vary by one order of magnitude at 50\% credible interval.  
$\mathcal{R}_{\rm{BBH}}(z)$ is inside the 90\% credible interval inferred from GWTC-2 (considering GW190814-like events) only if the metallicity spread is $\sigma_{Z} \lesssim 0.35$.

 By exploring 32 different models, we have varied only a small subset of all relevant model parameters, with sparse sampling of the many-dimensional space we considered. Hence, the effective uncertainty in the merger rate is likely higher than presented in our results. More exploration of the parameter space, and in particular of the $\alpha_{\rm CE}-$ natal kick space, is desirable in the future, even if it represents a computational challenge for population-synthesis models (e.g., \citealt{wong2019}).  
 
%\textbf{Future works of the authors may study the parameter space as a whole in order to suppress the systematic uncertainties that arises when varying a single parameter at a time and to identify the likely inter-correlation between parameters.} 

In summary, the uncertainties on both cosmic metallicity and binary evolution processes substantially affect the merger rate of BBHs and BHNSs. As shown in previous work \citep[e.g.,][]{rodriguezloeb2018,santoliquido2020,mapelli2020b}, dynamics in dense star clusters represents another important source of uncertainty for the BBH merger rate.

In contrast, BNSs are not much affected by metallicity evolution and are not dramatically influenced by dynamics either, because they are significantly less massive than BBHs \citep{ye2020,rastello2020,santoliquido2020}. Unlike BHs, for which the primordial BH formation channel has been proposed \citep{carr1974,carr2016}, BNSs can originate only from the death of massive stars. This set of lucky circumstances gives us the opportunity to use the BNS merger rate to put constraints on some extremely uncertain binary evolution processes, such as mass transfer, common envelope and natal kicks.

Our results already point to an intriguing direction: only 
large values of $\alpha_{\rm CE}$ ($\ge{}2$) and moderately low natal kicks (depending on the ejected mass and the SN mechanism) can match the cosmic merger rate inferred from GWTC-2.  
The growing sample of GW events will help us deciphering this puzzle.

%\textbf{A future development of this type of study would be to explore the parameter space as a whole in a multidimensional space, in order to assess and discard the inter-correlation among each parameter. Moreover, any comparison of our predicted models with the 90\% credible interval inferred by the LIGO-Virgo collaboration should take into account the systematic uncertainty arising from the other parameters while kept fixed. This key concept is well shown from Figure \ref{fig:err} where we have seen the propagation of the cosmological observational quantities on the merger rate density. This source of uncertainty must be always taken in account while varying the parameters that determine the binary evolution.}

\section*{Acknowledgements}
The authors thank the anonymous referee for their useful comments. We thank Michele Guadagnin, Alessandro Lambertini, Alice Pagano and Michele Puppin for useful discussions on mass transfer. MM, FS, NG and YB acknowledge financial support from the European Research Council for the ERC Consolidator grant DEMOBLACK, under contract no. 770017. MCA and MM acknowledge financial support from the Austrian National Science Foundation through FWF stand-alone grant P31154-N27.

%%%%%%%%%% DATA %%%%%%%%%%%%%%%%
\section*{Data availability}
The data underlying this article will be shared on reasonable request to the corresponding authors.

%\clearpage
\bibliographystyle{mnras}
\bibliography{santoliquido} 

%%%%%%%%%%%%%%%%%%%%%%%%%%%%%%%%%%%%%%%%%%%%%%%%%%
%%%%%%%%%%%%%%%%% APPENDICES %%%%%%%%%%%%%%%%%%%%%
%\clearpage
%\appendix
%\section{Mass spectrum}

% Don't change these lines
\bsp	% typesetting comment
\label{lastpage}
\end{document}